\documentclass[12pt]{article}

\usepackage[letterpaper]{geometry}  % for paper size and setting margin sizes
\usepackage[english]{babel}         % set language for the document
\usepackage{graphicx}               % for figures
\usepackage{float}                  % to use the "H" placement for figures
\usepackage[colorlinks=true, allcolors=blue]{hyperref}  % for hyperlinks
\usepackage{amsmath}
\usepackage{mathrsfs}
\usepackage{amsfonts}
\usepackage{csvsimple}
\usepackage{tabularx}
\usepackage{amsthm}
\usepackage{subcaption}
\usepackage{xcolor}
\usepackage{authblk}

\numberwithin{equation}{section}
\graphicspath{ {images/} } % specify the default path for images

\title{\bf\Large Polynomial deformations of $sl(2)$ and unified algebraic framework for solutions of a class of spin models}

\author[1]{Siyu Li \footnote{Siyu.Li@latrobe.edu.au}}
\author[2]{Ian Marquette \footnote{i.marquette@latrobe.edu.au}}
\author[3]{Yao-Zhong Zhang \footnote{yzz@maths.uq.edu.au}}

\affil[1]{\em Department of Mathematical and Physical Sciences, La Trobe University, Bundoora, VIC 3086, Australia}
\affil[2]{\em Department of Mathematical and Physical Sciences, La Trobe University, Bendigo, VIC 3552, Australia}
\affil[3]{\em School of Mathematics and Physics, The University of Queensland, Brisbane, QLD 4072, Australia}

\begin{document}

\maketitle

\begin{abstract}

\noindent We introduce novel polynomial deformations of the Lie algebra $sl(2)$. We construct their finite-dimensional irreducible representations and the corresponding differential operator realizations. We apply our results to a class of spin models with hidden polynomial algebra symmetry and obtain the closed-form expressions for their energies and wave functions by means of the Bethe ansatz method. The general framework enables us to give an unified algebraic and analytic treatment for three interesting spin models with hidden cubic algebra symmetry: the Lipkin-Meshkov-Glick (LMG) model, the molecular asymmetric rigid rotor, and the two-axis countertwisting squeezing model. 
We provide analytic and numerical insights into the structures of the roots of the Bethe ansatz equations (i.e. the so-called Bethe roots) of these models. We give descriptions of the roots on the spheres using the inverse stereographic projection. The changes in nature and pattern of the Bethe roots on the spheres indicate the existence of different phases of the models. We also present the fidelity and derivatives of the ground-state energies (with respect to model parameters) of the models. The results indicate the presence of critical points and phase transitions of the models. In the appendix, we show that, unlike the so-called Bender-Dunne polynomials, the set of polynomials in the energy $E$, $P_\ell(E)$,  corresponding to each of the three spin models has two critical polynomials whose zeros give the quasi-exact energy eigenvalues of the model. Such types of polynomials seem new.

\end{abstract}

\section{Introduction}

Polynomial deformations of Lie algebras are finitely generated non-linear algebras which 
lie at the intermediate place between Lie algebras and quantum algebras (which are quantum deformations of Lie algebras). 
Such structures occur as symmetry algebras in a variety of quantum mechanical systems. Polynomial algebras of degrees 2 and 3 are known as quadratic and cubic (or sometimes Higgs) algebras in the literature, respectively.  Lower-degree polynomial algebras were observed and used in the early 90's in the study of, e.g., nonlinear Yang-Mills theory \cite{schoutens1991}, integrable systems \cite{granovskii1992,freidel1991} and re-coupling of angular momenta \cite{granovskii1992, zhedanov1992}.
In the context of superintegrable systems, the underlying hidden quadratic algebra structures provided algebraic derivation of the exact energy spectra  and wavefunctions of the models (see e.g. \cite{hoque2015,liao2018}). Moreover, they can be used to describe spectrum degeneracies of the systems \cite{letourneau1995, bonatsos1994}.
%They are also connected with aspects of orthogonal polynomials and special functions such as $3j$- and $6j$-symbols \cite{zhedanov1988}.

Polynomial algebras have also found applications in analytically solving quasi-exactly solvable (QES) systems.
A system is said to be QES if only a finite number of its energy spectrum and the corresponding eigenfunctions can be obtained exactly by algebraic or analytic means, see e.g. recent review paper \cite{turbiner2016} on QES systems and the references therein. One-dimensional QES problems were introduced in \cite{turbiner1987} and were related to differential operator realizations of the finite-dimensional representations of (linear) Lie algebra $sl(2)$ \cite{turbiner1988}. 
In this context, the Hamiltonians of the QES models are given by quadratic expressions of the generators of $sl(2)$ and such models were said to have a hidden $sl(2)$ algebra symmetry. As the differential operators preserve certain finite-dimensional polynomial spaces \cite{gomez2007}, the models with hidden $sl(2)$ symmetry are QES and their closed form expressions for energy eigenvalues and wavefunctions can be derived by the Bethe ansatz method \cite{zhang2012}. Special deformations of $sl(2)$ were discussed in \cite{beckers1999} and applied to some QES cases such as the sextic oscillator model and the second-harmonic generation problem \cite{debergh2000}. In \cite{lee2010,lee2011,lee2011b}, more general polynomial algebras were introduced and applied to find closed-form solutions for classes of spin-boson models of interest in physics.

In this paper, we introduce a novel class of polynomial deformations of $sl(2)$, denoted $pl(sl(2))$, and construct their finite-dimensional representations and the corresponding single variable differential operator realizations in the space of monomials. The polynomial algebras $pl(sl(2))$ are hidden symmetry algebras underlying a class of spin models of interest in physics. This is seen by the fact that the Hamiltonians of the spin models admit algebraizations in terms of the generators of $pl(sl(2))$. Applying the differential representations of $pl(sl(2))$ and following the procedure in \cite{lee2010,lee2011b}, we obtain the closed-form expressions for the energy eigenvalues and the corresponding eigenfunctions of the Hamiltonian equations (time-independent Schr\"odinger equations) of the spin models. As examples of application for our general framework, 
we revisit three interesting spin models, the Lipkin-Meshkov-Glick (LMG) model \cite{lipkin1965}, the asymmetric rigid rotor  \cite{king1947}  and the two-axis countertwisting squeezing model \cite{kitagawa1991,kitagawa1993}, each of which will be shown to have a hidden cubic algebra (given by $psl(sl(2))$ of degree 3) symmetry. 
Analytic solutions of the LMG model were studied in \cite{pan1999}, while the authors in \cite{jarvis2008} presented results on the asymmetric rigid rotor in quasi-classical limit. See also \cite{lee2011b} for partial solutions of these two cases. The two-axis countertwisting squeezing model %was established to reduce standard quantum noise and has applications to interferometers in quantum optics \cite{kitagawa1993}. This model 
was solved analytically in \cite{pan2017a,pan2017b} by a different method. However, within the method in \cite{pan2017a,pan2017b} the integer and half-integer spin cases had to be treated separately. In this paper,  we provide a unified treatment for the integer and half-integer cases of the two-axis model based on its hidden cubic algebra symmetry. In the general framework established in this paper, the Hamiltonians of all three above-mentioned models can be expressed in terms of the differential operators which realize the finite-dimensional representations of certain cubic algebras. This enables us to solve the three models in a unified way using the Bethe ansatz approach \cite{zhang2012,zhang2013}. 

By combining the algebraic results with numerical simulations, we also develop a new way to study possible quantum phase transitions of these models. Critical points and quantum phase transitions have been studied widely in the literature for various systems. One of the common approaches is to analyze spin-coherent states and determine the energy surfaces with classical trajectories and the phase diagrams \cite{Octavio2006, Octavio2005, Romera2014}. Other approaches in the literature involve the analysis of the density states of a system in the thermodynamic limit, because the differences between the density states in each region of the phase diagram provide important information about phase transitions \cite{Ribeiro2007, Ribeiro2008, Links2015}. 
Here we present an approach based on the analytic results of energy eigenvalues and wavefunctions, determined from the roots of a set of algebraic Bethe ansatz equations. To provide evidence of quantum phase transitions, we map the roots of the Bethe ansatz equations to a Majorana sphere via inverse stereographic projection. Such Majorana representations (also called Majorana polynomials) can provide a geometric view of spin-$j$ states on a Bloch sphere \cite{Majorana1932,Chryssomalakos2018}. They were applied in \cite{Ribeiro2007, Ribeiro2008} to analyze the roots of the Majorana polynomials and the phase regions for the even sectors of the LMG model in the thermodynamical limit.

This paper is organized as follows. In section \ref{pl(sl2) and differential ops}, we introduce the polynomial algebra $pl(sl(2)$ and construct its finite-dimensional representations and the corresponding single-variable differential operator realization. In section \ref{Solns of general spin models}, we consider a class of spin models with a hidden $pl(sl(2)$ algebra symmetry. We present the algebraizations for the Hamiltonians of the spin models in terms of generators of $pl(sl(2))$. Applying the differential realization and the Bethe ansatz technique, we give closed-form solutions for the energy eigenvalues and eigenfunctions of the Schr\"odinger type equations of the models. In section \ref{Three spin models}, we apply the general framework in sections \ref{pl(sl2) and differential ops} and \ref{Solns of general spin models} to the three spin model with cubic algebra (polynomial algebra $pl(sl(2))$ of degree 3) symmetry and present explicit expressions for their energies and wavefunctions. We present analysis on (the explicit patterns of) 
We map the roots of the Bethe ansatz equations for these models onto unit spheres using the inverse stereographic projection technique and show the explicit patterns of the Bethe roots on the spheres. We also provide numeric analysis on the fidelity and the first- and second-order derivatives of the energies with respect to the model parameters. These numeric results indicate possible phase transitions in the spaces of model parameters. In section \ref{Conclusion}, we provide a summary of our results. In the Appendix, we show that corresponding to each of the three spin models there is a set of polynomials in the energy $E$ which has two critical polynomials. The quasi-exact energy eigenvalues of each model are given by the zeros of the two critical polynomials.

\section{Novel polynomial deformation of $sl(2)$ and differential operator realization}\label{pl(sl2) and differential ops}

In this section, we introduce a novel polynomial deformation, denoted $pl(sl(2))$, of the Lie algebra $sl(2)$. Throughout, we use $j$ (which is integer or half-integer) to label a spin-$j$ representation of $sl(2)$.  

Let us first recall some well-known results of the $sl(2)$ algebra. Let $J_\pm$, $J_0$ denote its generators. They satisfy the commutation relations
\begin{equation}\label{su(2) commutation relation}
    [J_0,J_{\pm}]=\pm J_{\pm},\qquad [J_+,J_-]=2 J_0. 
\end{equation}
The Casimir is 
\begin{equation}
    C=  J_+ J_- + J_0 (J_0-1) = J_- J_+ + (J_0+1) J_0. 
\end{equation}
Define the lowest weight state
\begin{equation}
   J_0 |j,0\rangle =-j |j,0\rangle, \quad \quad J_{-}|j,0\rangle =0,
\end{equation} 
where $j$ is non-negative integer or half integer. Then the general basis vectors in the irreducible representation space $V_j$, $j=0,\frac{1}{2}, 1, \ldots$, of $sl(2)$ are  
\begin{align}\label{su(2) j,m j,0}
    \begin{aligned}
        |j,m\rangle=\sqrt{\frac{(2j-m)!}{m!(2j)!}} J_+^m |j,0\rangle,
    \end{aligned}
\end{align}
and the action of $J_{0,\pm}$ on these vectors is given by
\begin{align}
    &\begin{aligned}\label{su(2) act on states J0}
        J_0 |j,m\rangle=(-j+m)|j,m\rangle,
    \end{aligned}\\
    &\begin{aligned}\label{su(2) act on states J+}
        J_+|j,m\rangle=\sqrt{(m+1)(2j-m)}|j,m+1\rangle,
    \end{aligned}\\
    &\begin{aligned}\label{su(2) act on states J-}
        J_-|j,m\rangle=\sqrt{m[j-(m-1)]}|j,m-1\rangle. 
    \end{aligned}
\end{align}
It is easily seen that $J_+|j, 2j\rangle=0$. Thus, $m=0,1,\ldots, 2j$ and the representation $V_j$ is $(2j+1)$ dimensional.

We now introduce $pl(sl(2))$. Throughout we let $k\geq 1$ be a fixed positive integer.  Set
\begin{equation}\label{pl(su(2)) generators}
    P_{\pm}=J_{\pm}^k ,\qquad P_0 = \frac{1}{k} J_0. 
\end{equation}
Then it can be easily checked that the operators (\ref{pl(su(2)) generators}) satisfy the following commutation relations 
\begin{equation}\label{pl(su(2)) commutation relations}
\begin{split}
     &[P_0,P_{\pm}]=\pm P_{\pm}, \\
     &[P_{+},P_{-}]= \phi^{(2k)}(P_0,C) - \phi^{(2k)}(P_0-1,C), 
\end{split}
\end{equation}
where $C$ is the Casimir operator of $sl(2)$  and $\phi^{(2k)}(P_0,c)$ is the polynomial of degree $2k$ in $P_0$ and $C$, 
\begin{equation}
 \phi^{(2k)}(P_0,C)=- \prod_{i=1}^{k} [C-( k P_0 +k -i+1)(k P_0 +k-i)]+ \prod_{i=1}^{k} [C- i (i-1)].  
\end{equation}
The algebra with commutation relations (\ref{pl(su(2)) commutation relations}) is a polynomial deformation of $sl(2)$, denoted $pl(sl(2))$, as these relations reduce to (\ref{su(2) commutation relation}) when $k=1$. Note that $pl(sl(2))$ is a polynomial algebra of degree $2k-1$.
The Casimir of $pl(sl(2))$ is given by
\begin{equation}
 K= P_{+}P_{-} + \phi^{(2k)}(P_0-1,C)= P_{-}P_{+} + \phi^{(2k)}(P_0 ,C). 
\end{equation}
% \[ = P_{-}P_{+} + \phi^{(2k)}(P_0 ,c)  \]
In the realization (\ref{pl(su(2)) generators}), the Casimir takes the particular form
\begin{equation} 
  K=  \prod_{i=1}^k [C-i(i-1)]. 
\end{equation}

We now construct the representations of $pl(sl(2))$ with the generators $P_{0,\pm}$ realized by (\ref{pl(su(2)) generators}). It can be easily checked that there are ${\rm min}(k, 2j+1)$ lowest weight states, 
%\begin{equation}\label{pl(su(2)) lowest states}
%     |j,0\rangle, \quad J_{+}|j,0\rangle,\quad...\;,\quad J_{+}^{k-1}|j,0\rangle
%\end{equation}
%with $2j \geq k-1$. We rewrite these lowest states as
\begin{equation}\label{pl(su(2)) lowest states}
    |j,0;p\rangle \sim J_+^p |j,0\rangle, \qquad p=0,1,\cdots, {\rm min}(k-1, 2j), 
\end{equation}
defined by $ P_- |j,0;p\rangle=0$. Here $|j,0\rangle$ is the lowest weight state of $sl(2)$. This implies that the irreducible $sl(2)$ module $V_j$ is decomposed into the direct sum of irreducible $pl(sl(2))$ modules $V_{j,p}$ as follows \cite{lee2011b}
\begin{align}
    &\begin{aligned}
        V_j = \bigoplus_{p=0}^{\text{min}(k-1,2j)} V_{j,p}, \qquad %k-1\leq 2j,\quad 
        \text{dim}V_j=2j+1. 
    \end{aligned}
\end{align}
It is obvious that $V_{j,p}$ is spanned by the states 
\begin{equation}\label{pl(su(2)) j,m,;p state}
    |j,m;p\rangle = \sqrt{\frac{(2j-p-km)!}{(p+km)!(2j)!}} J_{+}^{p+km}|j,0\rangle. 
\end{equation}
The action of $P_{0,\pm}$ on these states are given by
\begin{align}
&\begin{aligned}\label{pl(su(2)) P+ act on state}
  P_{+}|j,m;p\rangle = %\sqrt{\frac{(2j-p-km)!}{(p+km)!(2j)!}}  J_{+}^{p+k(m+1)}|j,0\rangle= 
  \prod_{i=1}^k \left[  (p+km+i)(2j -p  +1 -km -i)\right]^{\frac{1}{2}} |j,m+1;p\rangle ,
\end{aligned}\\
&\begin{aligned}\label{pl(su(2)) P- act on state}
 P_{-}|j,m;p\rangle = %\sqrt{\frac{(2j-p-km)!}{(p+km)!(2j)!}}  J_{-}^{k}J_{+}^{p+km}|j,0\rangle = 
 \prod_{i=1}^k \left[ (p+km-i+1)(2j-km-p+i)  \right]^{\frac{1}{2}} |j,m-1;p\rangle ,
\end{aligned}\\
&\begin{aligned}\label{pl(su(2)) P0 act on state}
        P_0|j,m;p\rangle=\left(m+\frac{p-j}{k}\right)|j,m;p\rangle. 
\end{aligned}
\end{align}

It can be shown that
\begin{equation}
    P_+ |j,{\cal N};p\rangle=0, \qquad {\cal N}\equiv \frac{2j-p-q}{k},
\end{equation}
where $q$ is a non-negative integer taking specific values $q = 0, 1, \ldots ,\text{min}(k -1, 2j)$ according to $j$ and $p$ so that ${\cal N}$ is always a non-negative integer. Such an $q$ always exists. This can be shown as follows. For any given $j$ and $p$ (corresponding to an irreducible representation labeled by $j$ and $p$), we can always write $2j-p=k\,{\cal N}+q'$ for some non-negative integer ${\cal N}$, with $q'$ being the reminder, i.e. $q'=0,1,\ldots,\text{min}(k-1,2j)$. Then we can choose $q=q'$ to cancel the reminder, as demonstrated in Table 1 below:
\begin{table}[ht]
    \centering
    \begin{tabular}{|c|c|c|c|c|c|}
     \hline
          & $p=0$ & $p=1$  & $p=2$  & .... & $p=k-1$   \\ 
        \hline
        $2j=k{\cal N}$   & $q=0$    & $q=k-1$  &  $q=k-2$  & ....  &  $q=1$  \\ 
        \hline
        $2j=k{\cal N}+1$   &  $q=1$   &  $q=0$   &  $q=k-1$  & ....  &  $q=2$ \\ 
        \hline
        $2j=k{\cal N}+2$       &  $q=2$    & $q=1$  & $q=0$   & ...  & $q=3$ \\ 
        \hline
        ...    &   ...  & ...   & ...  & ....  & ...  \\ 
        \hline
        $2j=k{\cal N}+k-1$    &   $q=k-1$  &  $q=k-2$  &  $q=k-3$   & ...  & $q=0$  \\ 
         \hline
    \end{tabular}
    \caption{Allowed values for $q$ corresponding to different $p$ and $j$. }
    \label{tab:my_label}
\end{table}
Thus, we have $m=0, 1, \ldots, {\cal N}$, and (\ref{pl(su(2)) P+ act on state})-(\ref{pl(su(2)) P0 act on state}) form a finite-dimensional representation of (\ref{pl(su(2)) commutation relations}) of dimension ${\cal N}+1$.

%In the following, we will denote the non-negative integer $\frac{2j-p-q}{k}$ by ${\cal N}$, i.e.
%\begin{equation}
%{\cal N}=\frac{2j-p-q}{k}.
%\end{equation}
The finite-dimensional irreducible representation of $pl(sl(2))$, (\ref{pl(su(2)) P+ act on state})-(\ref{pl(su(2)) P0 act on state}) can be realized by differential operators. Recall that $sl(2)$ can be realized by the single-variable differential operators in $z$ as
\begin{equation}\label{differential ops of sl2}
    J_+=z\left(-z\frac{d}{dz}+2j\right),\qquad J_-=\frac{d}{dz},\qquad J_0=z\frac{d}{dz}-j.
\end{equation}
Using the expression for $J_+$ and identifying $|j,0\rangle$ with $z^0$, (\ref{pl(su(2)) j,m,;p state}) is mapped to the monomial in $z$,
\begin{eqnarray}\label{pl(su(2)) jmp polynomial realization}
 |j,m;p\rangle &=& \sqrt{\frac{(2j-p-km)!}{(p+km)!(2j)!}} \left[z\left(-z\frac{d}{dz}+2j\right) \right]^{p+km} z^0\nonumber\\
 &=&\sqrt{\frac{(2j)!}{(p+km)!(2j-p-km)!}}\;z^{km+p}.
\end{eqnarray}
As $j$ and $k$ are fixed, we can remove the factor $(2j)!$ and set $x=z^k$ to simplify the monomial to the form,
\begin{equation}\label{untransformed basis}
  |j,m;p\rangle\quad \sim \quad %\frac{x^{m+p/k}}{\sqrt{(p+km)(2j-p-km)!}}=
  x^{p/k}\times \frac{x^{m}}{\sqrt{(p+km)(2j-p-km)!}}.
\end{equation}
The factor $x^{p/k}$ can be removed by a simple basis transformation (called gauge transformation) and so we will work with the transformed basis vectors 
\begin{equation}\label{transformed basis}
  \frac{x^{m}}{\sqrt{(p+km)(2j-p-km)!}}.
\end{equation}
 It can be shown that with respect to this transformed basis, the single-variable differential operator realization of the $({\cal N}+1)$-dimensional representation (\ref{pl(su(2)) P+ act on state})-(\ref{pl(su(2)) P0 act on state}) of $pl(sl(2))$ is given by
\begin{align}
    &\begin{aligned}\label{pl(su(2)) operator P+}
    P_+=x\prod_{i=1}^k\left(2j-p -i +1 -kx\frac{d}{dx}\right),
    \end{aligned}\\
    &\begin{aligned}\label{pl(su(2)) operator P-}
        P_{-}= x^{-1} \prod_{i=1}^k \left(p -i +1+k x \frac{d}{dx} \right),
    \end{aligned}\\
    &\begin{aligned}\label{pl(su(2)) operator P0}
        P_0 =x\,\frac{d}{dx}+\frac{p -j}{k}.
    \end{aligned}
\end{align}
Note that there are no $x^{-1}$ terms in $P_{-}$ since $\displaystyle \prod_{i=1}^k\, (p-i+1)=0$ for all the allowed values of $p$. Thus, differential operators $P_{\pm}, P_0$ are non-singular. 

Some remarks are in order. Firstly, note that $x=z^k$ above and throughout the remaining sections of the paper. Secondly, the above differential operator realization is for the ``gauged-transformed" generators $P_{\pm,0}$ with respect to the basis vectors (\ref{transformed basis}) of the space $\text{span}(1,x,x^2,\ldots, x^{\cal N})$. 

\section{Exact solutions to a class of spin models with hidden $pl(sl(2))$ polynomial algebra symmetry}\label{Solns of general spin models}

We consider a class of spin systems with the Hamiltonian of the form
\begin{equation}
    {\cal H}=c_+\,J_+^k+c_-\,J_-^k+\sum_{s=1}^k\,c_0\,J_0^s+c_*,
\end{equation}
where $c_+, c_-, c_s, c_*$ are model parameters so that ${\cal H}$ is hermitian. In terms of the generators of the degree-$(2k-1)$ polynomial algebra $pl(sl(2))$ the Hamiltonian can expressed as \begin{equation} \label{pl(su(2) Hamiltonian}
        {\cal H}=c_+\,P_++c_-\,P_-+\sum_{s=1}^k\,c_s\,\left(k\,P_0\right)^s+c_*. 
\end{equation}
 This means that the Hamiltonian admits a $pl(sl(2))$ algebraization and the corresponding spin model possesses a hidden $pl(sl(2))$ polynomial algebra symmetry. 
 
 The Hamiltonian equation of the system reads
 \begin{equation}
     {\cal H}|\psi\rangle=E|\psi\rangle.
 \end{equation}
 From (\ref{pl(su(2)) P+ act on state})-(\ref{pl(su(2)) P0 act on state}), $|\psi\rangle$ has the form $\displaystyle |\psi\rangle=\sum_{m=0}^{\cal N}\,a_m\,|j,m;p\rangle$, where $a_m$ are constant coefficients. We can equivalently write the Hamiltonian as a $k$-th order differential operator in the $x$-representation. Then the Hamiltonian equation becomes ${\cal H}\psi(x)=E\psi(x)$, where wavefunction $\psi(x)$ is given by a linear combination of basis vectors in (\ref{untransformed basis}). To use the differential operator realization (\ref{pl(su(2)) operator P+})-(\ref{pl(su(2)) operator P0}) of $P_{\pm,0}$ in the basis (\ref{transformed basis}), we make the gauge transformation of the wavefunction,
 \begin{equation}\label{general wave function}
     \psi(x)=x^{p/k}\,\varphi(x),\quad p=0,1,\ldots, k-1,
 \end{equation}
 where and from now on, for simplicity and without the loss of generality, we assume $k-1\leq 2j$ so that $\text{min}(k-1, 2j)=k-1$. After this transformation, the Hamiltonian equation becomes
 \begin{eqnarray}\label{Hamiltonian eq}
    H\varphi(x)=E\varphi(x), 
 \end{eqnarray}
where $H$ is the gauged-transformed $k$-th order differential operator given by 
\begin{eqnarray}\label{gauge-transformed H}
    H&=&x^{-p/k}\,{\cal H}\,x^{p/k}\nonumber\\
     &=&c_+\,x\prod_{i=1}^k\left(2j-p -i +1 -kx\frac{d}{dx}\right) %\nonumber\\
     +c_-\,x^{-1} \prod_{i=1}^k \left(p -i +1+k x \frac{d}{dx} \right)  \nonumber\\
     & &+\sum_{s=1}^k\,c_s\left(kx\,\frac{d}{dx}+{p -j}\right)^s+c_*, \qquad p=0,1,\ldots, k-1.
\end{eqnarray}
Here, the differential operator realization (\ref{pl(su(2)) operator P+})-(\ref{pl(su(2)) operator P0}) for the ``gauged-transformed" $P_{\pm,0}$ has been used.

Analytic solutions of the Hamiltonian equation,  $ H\varphi(x)=E\varphi(x)$, can be obtained following the procedure used in \cite{lee2011b,lee2010,lee2011}. It is easy to verify that for any $n\in\mathbb{Z}_+$
\begin{eqnarray}
    Hx^n=x^{n+1}\prod_{i=1}^k\left(2j-p-i+1-kn\right)+\text{lower order terms}.
\end{eqnarray}
When $n={\cal N}$, the first term on the right-hand side of the above equation disappears. That is, $H$ preserves an invariant polynomial subspace of degree ${\cal N}$, $H{\cal V}\subseteq {\cal V},\; {\cal V}=\text{span}(1,x,\ldots,x^{\cal N})$ and thus is quasi-exactly solvable. It follows that the Hamiltonian equation (\ref{Hamiltonian eq}) admits closed-form, polynomial solution of the form
\begin{equation}\label{wavefunction in pl(su(2)) BA}
    \varphi(x)=\prod_{i=1}^{\cal N}\,(x-x_i),\qquad \varphi(x)\equiv 1 ~\text{for}~{\cal N}=0,
\end{equation}
where $\{x_i|i=1,2,\ldots,{\cal N}\}$ are the roots of the polynomial to be determined below.

We rewrite the Hamiltonian (\ref{gauge-transformed H}) as
\begin{equation}\label{Hamiltonian in pl(su(2)) polynomial realization}
    H=\sum_{i=1}^k P_{i}(x) \frac{d^i}{dx^i}+P_0(x). 
\end{equation}
Where $P_{i}(x)$ and $ P_0(x)$ are polynomials in $x$ obtained from the expansion of the products in the differential operator realization. Substituting (\ref{wavefunction in pl(su(2)) BA}) into (\ref{Hamiltonian in pl(su(2)) polynomial realization}), we obtain
\begin{equation}\label{Energy in pl(su(2))}
    E=\frac{H \varphi}{\varphi}=\sum_{i=1}^k P_{i}(x)\,i!\,\sum_{\alpha_1<\alpha_2<...\alpha_i}^{\cal N}\frac{ i!}{(x-x_{\alpha_1})(x-x_{\alpha_2})...(x-x_{\alpha_i})}+P_0(x).
\end{equation}
The l.h.s of (\ref{Energy in pl(su(2))}) is constant, while the r.h.s is a meromorphic function in $x$ with at most single poles. For them to be equal, we need to eliminate all singularities on the r.h.s. of the equation. This can be achieved by demanding all the residues of the simple poles at $x=x_{i},\;i=1,2,\ldots,{\cal N}$, equal to 0 \cite{lee2010,lee2011}. This leads to the algebraic equations for the roots $\{x_i\}$,
\begin{equation}\label{BAE for polynomial algebra}
    \sum_{i=2}^k  \sum_{\alpha_1<\alpha_2<...\alpha_{i-1}\neq \alpha}^{\cal N}\frac{P_{i}(x_\alpha)\, i!}{(x_{\alpha}-x_{\alpha_1})(x_{\alpha}-x_{\alpha_2})...(x_{\alpha}-x_{\alpha_{i-1}})}+P_1(x_{\alpha})=0, \quad \alpha=1,2,\ldots, {\cal N}. 
\end{equation}
(\ref{BAE for polynomial algebra}) are called Bethe ansatz equations. 

We now determine the corresponding eigenvalue $E$ of the Hamiltonian equation. It is easy to check that solution (\ref{wavefunction in pl(su(2)) BA}) has the leading-order expansion,
\begin{equation}\label{leading order of psi}
    \varphi(x)=x^{\cal N}-x^{{\cal N}-1}\sum_{i=1}^{\cal N}\,x_i+\cdots,
\end{equation}
Accordingly, the leading-order expansions for $P_{\pm,0}\,\psi(x)$ are given by
\begin{eqnarray}\label{leading order of p+-0 psi}
   P_+\,\varphi(x)&=&-x^{\cal N}\,\left[\prod_{i=1}^k\left(2j-p-i+1-k({\cal N}-1)\right)\right]\sum_{i=1}^{\cal N}\,x_i+\cdots,\nonumber\\
   P_-\,\varphi(x)&=&x^{{\cal N}-1}\,\prod_{i=1}^{\cal N}(p-i+1+k{\cal N})+\cdots,\nonumber\\
   P_0\,\varphi(x)&=&x^{\cal N}\,\left({\cal N}+\frac{p-j}{k}\right)+\cdots.
\end{eqnarray}
Substituting (\ref{leading order of p+-0 psi}) into the Hamiltonian equation and equating the $x^{\cal N}$ terms, we obtain the quasi-exact energy eigenvalue
\begin{eqnarray}
    E=c_*+\sum_{s=1}^k\,c_s\,\left(k{\cal N}+{p-j}\right)^s-c_+\,\left[\prod_{i=1}^k\left(2j-p-i+1-k({\cal N}-1)\right)\right]\sum_{i=1}^{\cal N}\,x_i,
\end{eqnarray}
where $x_i$ satisfy the Bethe ansatz equations 
(\ref{BAE for polynomial algebra}). 

The corresponding wave function of the system is
\begin{equation}
\psi(x)=x^{p/k}\,\varphi(x)=x^{p/k}\,\prod_{i=1}^{\cal N}(x-x_i),\qquad p=0,1,\ldots, k-1.
\end{equation}
In terms of the variable $z$ (recalling $x=z^k$), the wave function takes the form
\begin{equation}
\psi(z)=z^{p}\,\varphi(z)=z^{p}\,\prod_{i=1}^{\cal N}(z^k-x_i),\qquad p=0,1,\ldots, k-1.
\end{equation}
Here $\{x_i\}$ are determined from the Bethe ansatz equations. Note that the values of $p$ and $q$ are constrained such that $\displaystyle {\cal N}=\frac{2j-p-q}{k}$ is a non-negative integer.

\section{Explicit examples: unified treatment of three spin models with cubic algebra symmetry}\label{Three spin models}

In this section, we revisit several well-known spin models whose Hamiltonians admit algebraizaitions in terms of the generators of the cubic deformations of $sl(2)$ (i.e. polynomial $pl(sl(2))$ algebra of degree 3). We will apply the general results of the previous section to obtain closed-form solutions of these models in a unified manner.

\subsection{LMG model}\label{LMG}

The Hamiltonian of the LMG model is 
\begin{equation}\label{LMG hamiltonian}
{\cal H}=\Delta\,  J_0 + g\, ( J_{+}^2 + J_{-}^2).
\end{equation}
This Hamiltonian admits an algebraization in terms of $pl(sl(2))$ of degree 3 corresponding to the $k=2$ case of the general results in the preceding section.  It follows that the LMG model Hamiltonian has a cubic algebra as its hidden symmetry. 

\subsubsection{Closed-form solutions}
Specializing the general results in the preceding section to (\ref{LMG hamiltonian}), we have $p=0,1$ and
\begin{eqnarray}\label{N for LMG}
    {\cal N}=j-\frac{p+q}{2}
\end{eqnarray}
with $q=0,1$ so that ${\cal N}$ is a non-negative integer. In terms of the differential realization of the cubic algebra, the gauge-transformed version $H$ of the Hamiltonian (\ref{LMG hamiltonian})  takes the form of a 2nd-order differential operator, and the corresponding Hamiltonian equation is nothing more than the Schr\"odinger equation. Explicitly,
\begin{equation}
\begin{split}
       &4g(x^3+x)\frac{d^2\varphi}{dx^2}+[2g(1+2p)+2\Delta x+2g(3-4j+2p)x^2]\frac{d\varphi}{dx}\\
       &\qquad +\left\{[2gj(2j-1)+gp(1-4j+p)]x-[E+(j-p)\Delta]\right\}\varphi=0. 
\end{split}
\end{equation}
Solutions of this equation is given by 
\begin{equation}
    \varphi(x)=\prod_{i=1}^{\cal N} (x-x_i), \qquad \varphi(x)\equiv 1 ~\text{for}~{\cal N}=0,
\end{equation}
where $\{x_i\}$ are the roots of Bethe ansatz equations
\begin{equation}\label{LMG BAE}
        \sum_{\ell\neq i}^{\cal N} \frac{2}{x_\ell-x_i}+\frac{g(3-4j+2p)x_i^2+\Delta x_i+g(1+2p)}{2g (x_i^3+x_i)}=0, \quad i=1,2,\cdots,{\cal N}. 
\end{equation}
The corresponding energy eigenvalue $E$ is 
 \begin{equation}
   E=(2{\cal N}+p-j)\,\Delta -g(2j-p+1-2{\cal N})(2j-p+2-2{\cal N})\sum_{i=1}^{\cal N}\,x_i. 
 \end{equation}
From the constraint (\ref{N for LMG}), 
\begin{equation}
    j={\cal N}+\frac{p+q}{2},\quad {\cal N}\in\mathbb{Z}_+; \quad p=0,1;\quad q=0,1. 
\end{equation}
This gives rise to the following four cases:
\begin{eqnarray}
    &&j~\text{even}:\quad p=q=0 \quad\text{or}\quad p=1=q,\\
    &&j~\text{odd}:\quad p=0,~ q=1\quad\text{or}\quad p=1,~ q=0,
\end{eqnarray}

We now include the gauge factor and apply $x=z^k$ to obtain the wavefunction in terms of $z$ for the LMG model,
\begin{equation}
    \psi(z)=z^p\prod_{i=1}^{\cal N} (z^2-x_i), \qquad p=0,1,
\end{equation}
where $p=0$ and $p=1$ correspond to even and odd sectors, respectively. We remark that the
wavefunction has the form of the Majorana-type polynomial related to the spin-$j$ representation of $sl(2)$ \cite{Majorana1932}
\begin{equation}
\begin{split}
    \psi(z)&=%\sum_{t=0}^{2j}\sqrt{\binom{2j}{t}} c_{t} z^t=
        \sum_{\ell=0}^{{\cal N}} \sqrt{\frac{(2j)!}{(p+k\ell)!(2j-p-k\ell)!}}\,A_{k\ell+p} \,z^{k\ell+p},
\end{split}
\end{equation}
where $A_{k\ell+p}$ are expansion coefficients, and  $p=0,1$ label the even and odd sectors of the wavefunction, respectively. Explicitly,
\begin{align}
    &\begin{aligned}\label{pl(su(2)) wavefunction even}
\psi(z)_{even}=\left\{\begin{array}{ll}
\displaystyle \sum_{\ell=0}^{j}\sqrt{\frac{(2j)!}{(2\ell)!(2j-2\ell)!}}\,A_{2\ell}\, z^{2\ell},&\quad \text{for}~j\in \mathbb{Z}_+,\\
& \\
\displaystyle \sum_{\ell=0}^{j-1/2}\sqrt{\frac{(2j)!}{(2\ell)!(2j-2\ell)!}}\,A_{2\ell} \,z^{2\ell},&\quad \text{for}~j\in \frac{1}{2}\mathbb{Z}_+,\\
\end{array}\right.
    \end{aligned}\\
    &\begin{aligned}\label{pl(su(2)) wavefunction odd}
 \psi(z)_{odd}=\left\{\begin{array}{ll}
\displaystyle \sum_{\ell=0}^{j-1}\sqrt{\frac{(2j)!}{(2\ell+1)!(2j-2\ell-1)!}}\,A_{2\ell+1} \,z^{2\ell+1},&\quad \text{for}~j\in \mathbb{Z}_+,\\
& \\
\displaystyle \sum_{\ell=0}^{j-1/2}\sqrt{\frac{(2j)!}{(2\ell+1)!(2j-2\ell-1)!}}\,A_{2\ell+1} \,z^{2\ell+1},&\quad \text{for}~j\in \frac{1}{2}\mathbb{Z}_+.\\
\end{array}\right.
    \end{aligned}
\end{align}

As examples, we consider ${\cal N}=0$ and ${\cal N}=1$ cases and present the corresponding energies and wavefunctions in the even and odd sectors of the model. 

For ${\cal N}=0$, the energy and wavefunction are given by $E=(p-j)\,\Delta$ and $\psi(z)=z^p$, respectively. Here $p$ (which takes the value 0 or 1) and the spin $j$ are constrained so that ${\cal N}=j-(p+q)/2$, where $q=0,1$, is equal to 0. In this case, the energies and wavefunctions in the even sectors ($p=0$) are 
\begin{equation}
    E_{even}=0,\qquad \psi_{even}(z)=1, 
\end{equation}
\begin{equation}
 E_{even}=-\frac{\Delta}{2},\qquad \psi_{even}(z)=1, 
\end{equation}
which correspond to $j=0$ (integer) and $j=1/2$ (half-integer), respectively.
The energies and wavefunctions in the odd sector ($p=1$) are given by
\begin{equation}
    E_{odd}=0,\qquad \psi_{odd}(z)=z,
\end{equation}
\begin{equation}
 E_{odd}=\frac{\Delta}{2},\qquad \psi_{odd}(z)=z, 
\end{equation}
associated with $j=1$ (integer) and $j=1/2$ (half-integer), respectively.

For ${\cal N}=1$, there are two sets of energies and wavefunctions in the the even sector ($p=0$), 
\begin{equation}
    E_{even}=\mp \sqrt{\Delta^2+4g^2},\qquad \psi_{even}(z)=z^2-\frac{\Delta \pm \sqrt{\Delta^2+4g^2}}{2g}, 
\end{equation}
\begin{equation}
 E_{even}=-\frac{\Delta}{2}\mp \sqrt{\Delta^2+12g^2},\qquad \psi_{even}(z)=z^2-\frac{\Delta \pm \sqrt{\Delta^2+12g^2}}{6g} 
\end{equation}
which correspond to $j=1$ (integer) and $j=3/2$ (half-integer), respectively.
Similarly, the energies and wavefunctions in the odd sector ($p=1$) are given by
\begin{equation}
    E_{odd}=\mp \sqrt{\Delta^2+36g^2},\qquad \psi_{odd}(z)=z^3-\frac{\Delta \pm \sqrt{\Delta^2+36g^2}}{6g}z,
\end{equation}
\begin{equation}
 E_{odd}=\frac{\Delta}{2}\mp \sqrt{\Delta^2+12g^2},\qquad \psi_{odd}(z)=z^3-\frac{\Delta \pm \sqrt{\Delta^2+12g^2}}{2g}z, 
\end{equation}
associated with $j=2$ (integer) and $j=3/2$ (half-integer), respectively. (We remark that other energies and wavefunctions for $j=2$ are obtained from ${\cal N}=2$ solutions.)

For higher values of ${\cal N}$,  the Bethe roots of the Bethe ansatz equations are difficult to determine analytically, and numerical techniques are usually needed in order to determine the roots. 

\subsubsection{Numerical analysis}

In this subsection, we present numerical results on the roots of the Bethe ansatz equations for higher values of ${\cal N}$. 
To look for evidence of possible phase transitions in this model, we will also examine the derivatives of the energy with respect to the model parameter $g$ and the fidelity of the lowest energy state.

In general, as can be seen from (\ref{pl(su(2)) wavefunction even}) and (\ref{pl(su(2)) wavefunction even}), a spin-$j$ state has $2j+1$ Bethe roots (or $2j+1$ Majorana zeros), so that $2j+1$ Majorana constellations appear on a 2-sphere for the corresponding solution. The Bethe roots in the even and odd sectors are determined by the Bethe ansatz equations. The energy for a spin-$j$ state can also be written in terms of Bethe roots.
To provide a visualization of spin-$j$ states, we map the roots of the polynomial solution onto a unit sphere by inverse stereographic projection, which enables us to analyze the patterns of the Bethe root distributions. 

\begin{figure}[h]
  \centering
  \begin{subfigure}{0.25\linewidth}
    \centering
    \includegraphics[width=\linewidth]{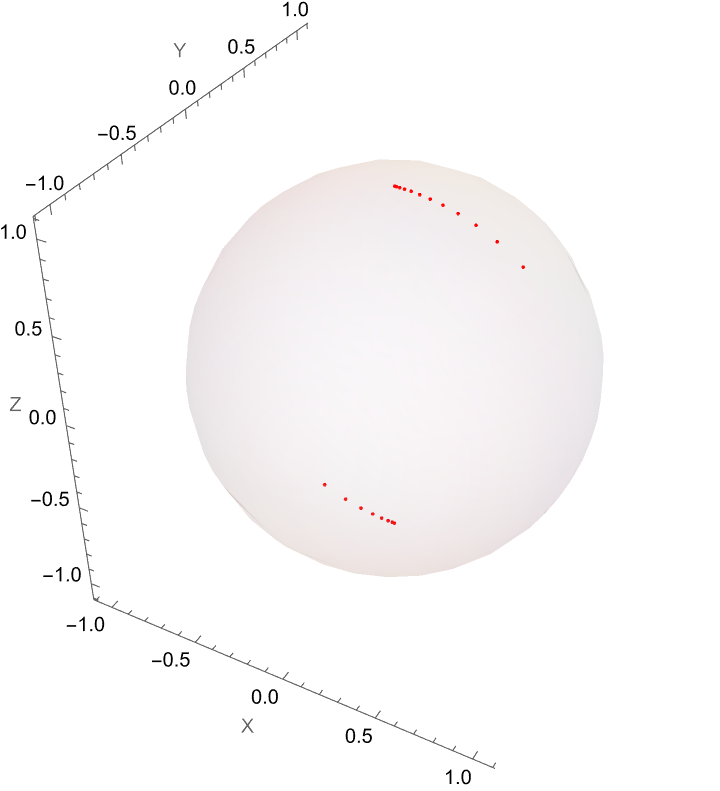}
    \caption{Root distribution for the lowest energy state on the unit sphere.}
    \label{f g=0.05 gs}
  \end{subfigure}
  \hfill
  \begin{subfigure}{0.25\linewidth}
    \centering
    \includegraphics[width=\linewidth]{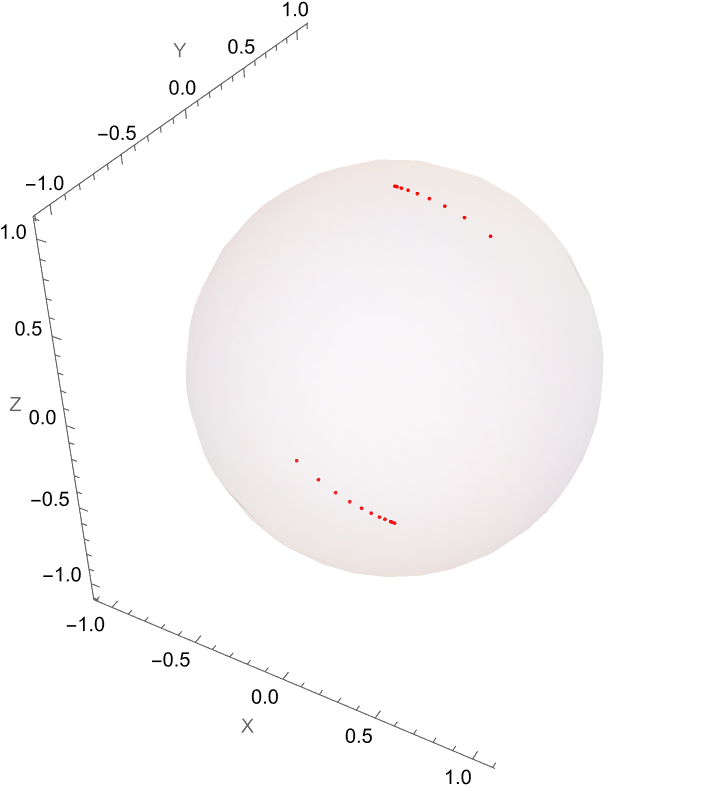}
    \caption{Roots start to form two arcs and a closed loop.}
    \label{f g=0.05 s2}
  \end{subfigure}
    \hfill
  \begin{subfigure}{0.25\linewidth}
    \centering
    \includegraphics[width=\linewidth]{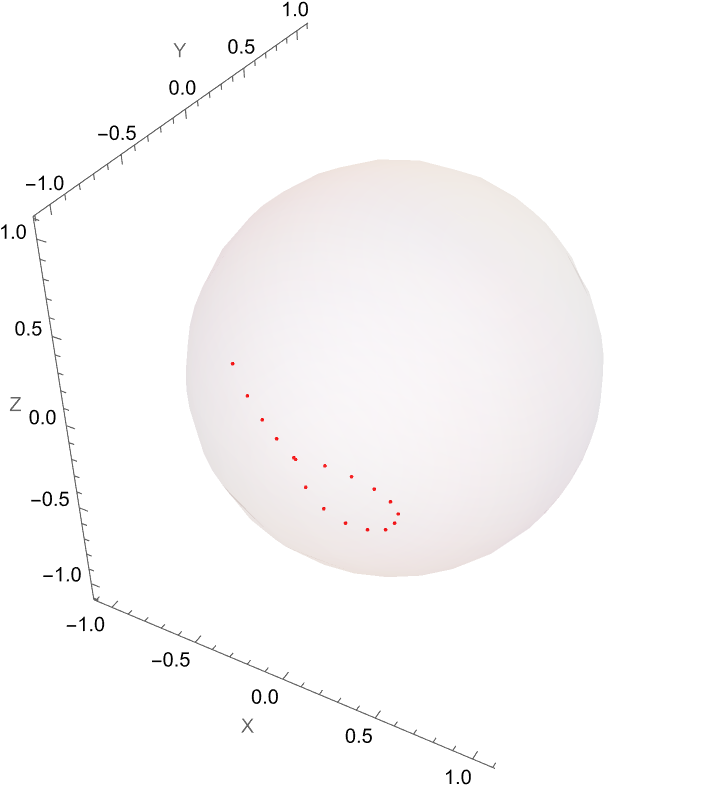}
    \caption{The closed loop expands to connect to one arc.}
    \label{f g=0.05 s3}
  \end{subfigure}
  \caption{Majorana representation for $g=0.05$ and change in root pattern with the change of model parameter.}
  \label{fig g=0.05}
\end{figure}

\begin{figure}[h]
  \centering
  \begin{subfigure}{0.25\linewidth}
    \centering
    \includegraphics[width=\linewidth]{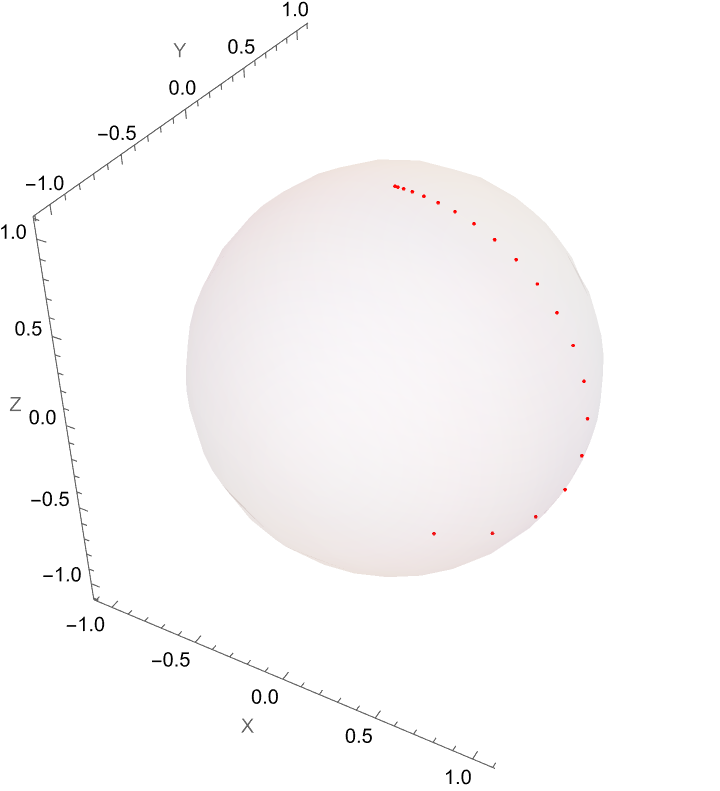}
    \caption{Bethe root distribution for the lowest energy state on the unit sphere. %This is different behavior as the parameter $g$ was changed
    }    \label{f g=0.15 gs}
  \end{subfigure}
  \hfill
  \begin{subfigure}{0.25\linewidth}
    \centering
    \includegraphics[width=\linewidth]{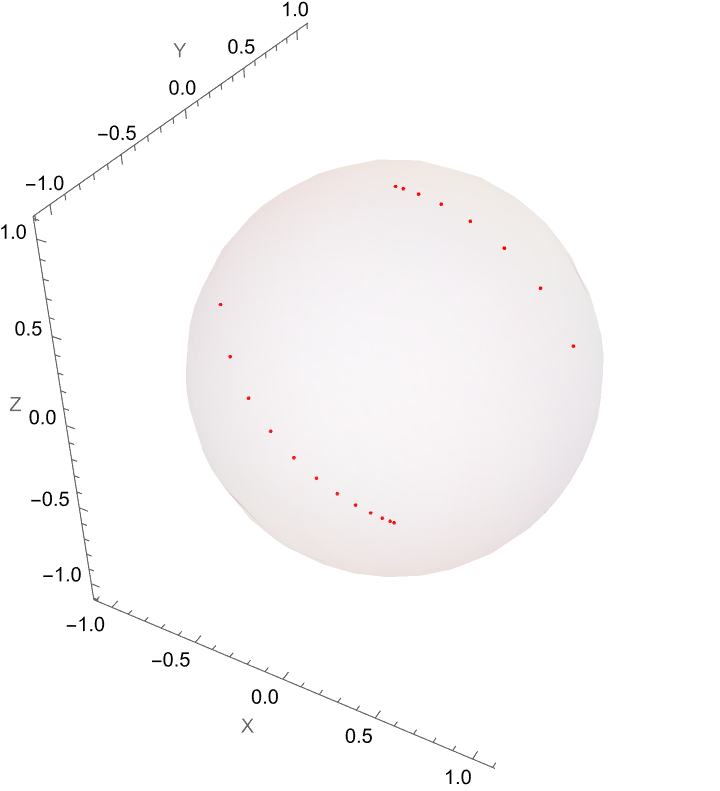}
    \caption{Root root density decreases to form two arcs, which is different from the $g=0.05$ case.
    }    \label{f g=0.15 s2}
  \end{subfigure}
    \hfill
  \begin{subfigure}{0.25\linewidth}
    \centering
    \includegraphics[width=\linewidth]{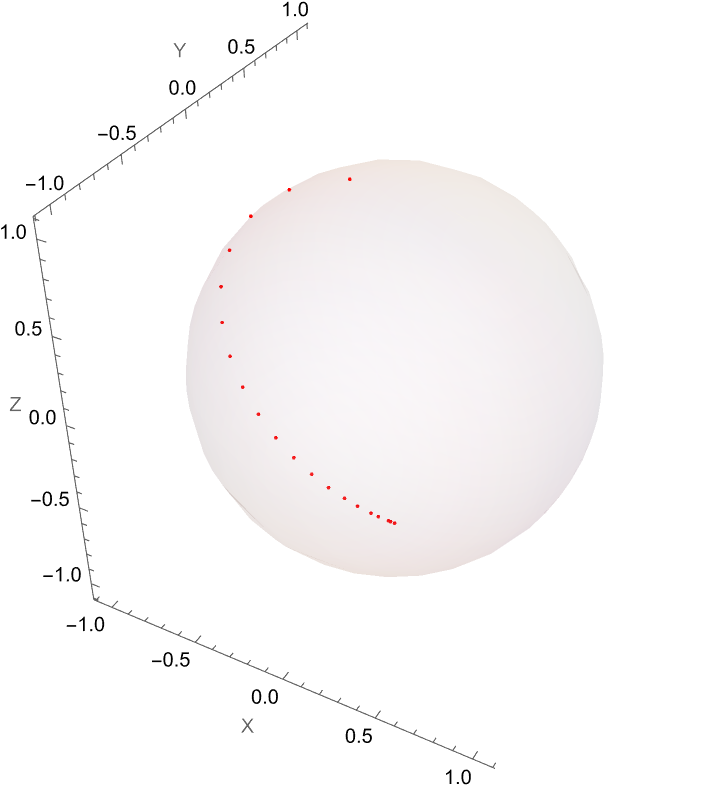}
    \caption{Root density on the other side of the unit sphere increases to form an open arc. %Again, this case differ from $g=0.05$
    }    \label{f g=0.15 s3}
  \end{subfigure}
  \caption{Majorana representation for $g=0.15$ and change in root pattern with the change of model parameter.}
  \label{fig g=0.15}
\end{figure}

There are two free parameters, $g$ and $\Delta$, in the Hamiltonian of the LMG model. We focus on the case of $j=20={\cal N}$, and set $\Delta=10-g$ and change $g$ from 0.05 to 0.6. Fig.\ref{fig g=0.05} and Fig. \ref{fig g=0.15} below provide the roots of the Bethe ansatz equations (\ref{LMG BAE}) mapped onto the unit sphere. 

%\newpage
\noindent Fig.\ref{f g=0.05 gs} is the root distribution for the lowest energy state with $g=0.05$. The roots separate into two parts and are located at both poles. The number of roots at the north pole decreases. Close to the south pole, the roots form a circle as seen in Fig.\ref{f g=0.05 s3}.

%\newpage
\noindent Fig.\ref{f g=0.15 gs} is the root distribution for the lowest energy state with $g=0.15$. All roots spread on the positive side of the $x$-axis in the complex plane, and with increasing energy they move to the negative side as seen in Fig.\ref{f g=0.15 s2} and Fig.\ref{f g=0.15 s3}. 
It is obvious that the root distribution varies with $g$.  The change in patterns in the Bethe root distributions indicates the presence of a certain phase transition. 

\begin{figure}[h]
  \centering
  \begin{subfigure}{0.56\linewidth}
    \centering
    \includegraphics[width=\linewidth]{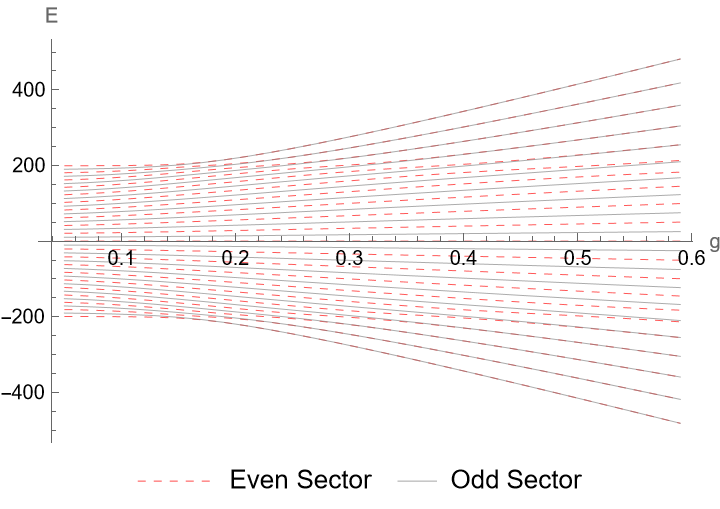}
    \caption{Two distinct regions for the energy spectrum. The concordance of the continuous and dash lines indicates degeneracies in the energy levels.}
    \label{f energy LMG}
  \end{subfigure}
  \hfill
  \begin{subfigure}{0.36\linewidth}
    \centering
    \includegraphics[width=\linewidth]{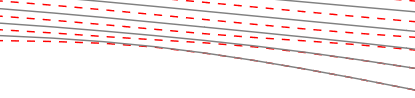}
    \caption{A zoom-in plot showing the change in the degeneracies of the energy levels.}
    \label{f energy LMG zoom}
  \end{subfigure}
  \caption{Energy spectrum of the LMG model}
  \label{fig LMG}
\end{figure}

%\newpage
\noindent Fig.\ref{f energy LMG} presents the energy spectrum of the LMG model in the odd and even sectors. The red dashed lines and the gray continuous lines correspond to the energies in the even and odd sectors, respectively. When $g$ is less than 1.8 approximately, the energies in both sectors are non-degenerate. When $g$ is greater than approximately 1.8, some lower-energy states and some higher-energy states appear to have energy degeneracies. Fig.\ref{f energy LMG zoom} is the zoom in for some low-energy states around $g=1.8$. Fig.\ref{fig LMG} indicates a possible phase transition appearing in $g=1.8$. The nondegenerate energies in the even and odd sectors lie in one phase region, and the degenerate ones are in another phase region. 

\begin{figure}[h]
  \centering
  \begin{subfigure}{0.32\linewidth}
    \centering
    \includegraphics[width=\linewidth]{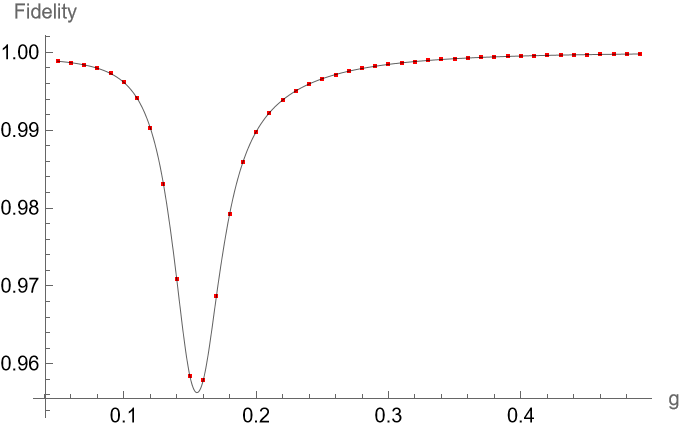}
    \caption{Fidelity calculated from the lowest energy state.}
    \label{f fidelity LMG}
  \end{subfigure}
  \hfill
  \begin{subfigure}{0.32\linewidth}
    \centering
    \includegraphics[width=\linewidth]{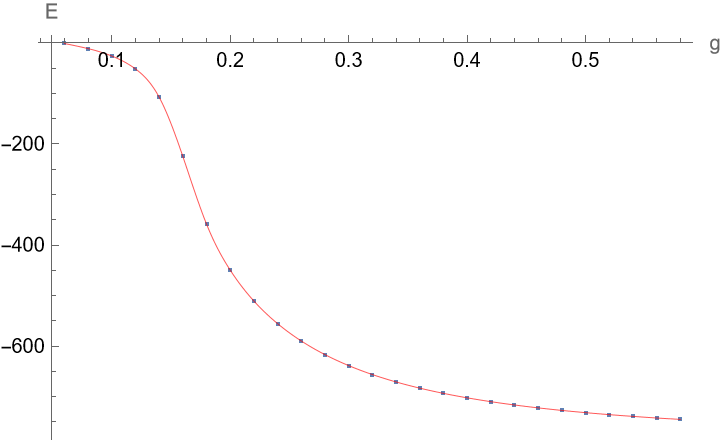}
    \caption{1st order derivative of the energy with respect to $g$.}
    \label{f LMG 1st derivative}
  \end{subfigure}
    \hfill
  \begin{subfigure}{0.32\linewidth}
    \centering
    \includegraphics[width=\linewidth]{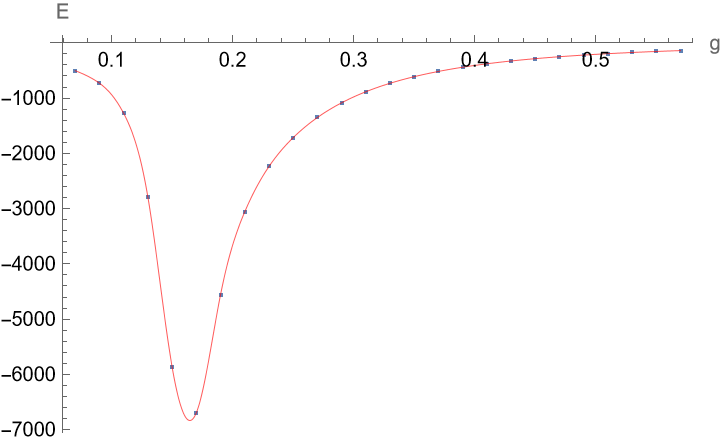}
    \caption{2nd order derivative of energy with respect to $g$.}
    \label{f LMG 2nd derivative}
  \end{subfigure}
  \caption{Fidelity and the first- and second-order derivatives of the energy.}
  \label{fig LMG others}
\end{figure}

%\newpage
\noindent Fig.\ref{fig LMG others} describes the physical properties of the lowest energy state.  Fig.\ref{f fidelity LMG} is a plot of fidelity %$F=|\langle \psi(g)|\psi(g+\delta g)\rangle|^2$, 
for the lowest energy state. The parameter value corresponding to the minimum fidelity is $g=1.8$, which is the value for the critical point and where a possible phase transition occurs. To see the change in energy during a phase transition, we plot the first- and second-order derivatives of the energy with respect to the model parameter $g$ in Fig.\ref{f LMG 1st derivative} and Fig.\ref{f LMG 2nd derivative}. The peak of the second-order derivative of the energy and the minimum of fidelity are evidence of different regions and phase transition at $g=1.8$ in the LMG model. 
 
Fidelity has been used in many instances to characterize different regions in a phase diagram and phase transitions for models in the context of condensed matter. Our analysis of fidelity is based on the exact polynomial solutions of the LMG model derived using the hidden cubic algebra structure.

\subsection{Molecular asymmetric rigid rotor}\label{Rotor}

The Hamiltonian of molecular asymmetric rigid rotor is  
\begin{eqnarray}\label{Asymmetric rotor hamiltonian}
   {\cal H}&=&a J_x^2 +b J_y^2 + c J_z^2\nonumber\\
   &=& \frac{2c-a-b}{2} J_0^2 + \frac{a-b}{4} (J_{+}^2 + J_{-}^2) + \frac{a+b}{2} C, \qquad a\neq b,
\end{eqnarray}
where $C$ is the Casimir of $sl(2)$ which takes the value $j(j+1)$ in a spin-$j$ representation.
This Hamiltonian has an algebraization in terms of $pl(sl(2))$ of degree 3 corresponding to $k=2$ in (\ref {pl(su(2)) commutation relations}). That is, it has a hidden cubic algebra symmetry. Note that the Hamiltonian of the asymmetric rigid rotor has the same form as that of the LMG model if one omits the term proportional to the Casimir operator $C$ which only shifts the origin of energy.

\subsubsection{Closed-form solutions}
Applying the general framework to the asymmetric rigid rotor Hamiltonian (\ref{Asymmetric rotor hamiltonian}), we find  $p=0,1$ and
\begin{eqnarray}\label{N for asymmetric rotor}
    {\cal N}=j-\frac{p+q}{2}
\end{eqnarray}
with $q=0,1$ so that ${\cal N}$ is a non-negative integer. In terms of the differential realization of the cubic algebra, the corresponding gauge-transformed Hamiltonian $H$ takes the form of 2nd-order differential operator and the Hamiltonian equation for $H$ is the Schr\"odinger equation which can be written as the form 
\begin{equation}
\begin{split}
        &[(a-b)x^3-2(a+b-2c)x^2+(a-b)x]\frac{d^2\varphi}{dx^2}\\
        &+\left[\frac{1}{2}(a-b)(3-4j+2p)x^2+2(a+b-2c)(j-p-1)x+\frac{1}{2}(a-b)(1+2p)\right]\frac{d\varphi}{dx}\\
        &+\left[\frac{1}{4}(a-b)(p-2j)(1-2j+p)x+\frac{1}{2}[j(a+b)(1+j)-2E-(a+b-2c)(j-p)^2]\right]\varphi=0. 
\end{split}
\end{equation}
Solutions of this equation are given by polynomials of degree ${\cal N}$,
\begin{equation}
    \varphi(x)=\prod_{i=1}^{\cal N} (x-x_i), \qquad \varphi(x)\equiv 1~\text{for}~{\cal N}=0,
\end{equation}
where $\{x_i\}$ are roots of the Bethe ansatz equations
\begin{eqnarray}\label{asymmetry top molecule BAE}
       \sum_{\ell\neq i}^{\cal N} \frac{2}{x_\ell-x_i}
        +\frac{(a-b)(3-4j+2p)x_i^2-4(a+b-2c)(1-j+p)x_i+(a-b)(1+2p)}{2(a-b) (x_i^3+x_i)-4(a+b-2c)x_i^2}=0,\nonumber\\
\end{eqnarray}
where $i=1,2,\cdots,{\cal N}$.
The corresponding energy eigenvalue $E$ is 
 \begin{eqnarray}\label{Energy for asymmetric rotor}
   E&=&\frac{a+b}{2}\,j(j+1)+\frac{2c-a-b}{2}\,(2{\cal N}+p-j)^2\nonumber\\
   & &-\frac{a-b}{4}\,(2j-p+1-2{\cal N})(2j-p+2-2{\cal N})\,\sum_{i=1}^{\cal N}\,x_i. 
 \end{eqnarray}
Similar to the LMG model case, from the constraint (\ref{N for asymmetric rotor}), 
\begin{equation}
    j={\cal N}+\frac{p+q}{2},\quad {\cal N}\in\mathbb{Z}_+; \quad p=0,1;\quad q=0,1, 
\end{equation}
we have the following four cases for the values of $j, p$ and $q$ so that ${\cal N}$ is non-negative integer:
\begin{eqnarray}
    &&j~\text{even}:\quad p=q=0 \quad\text{or}\quad p=1=q,\\
    &&j~\text{odd}:\quad p=0,~ q=1\quad\text{or}\quad p=1,~ q=0,
\end{eqnarray}

We now include the gauge factor and apply $x=z^2$ to obtain the wavefunction in terms of $z$ for the asymmetric rigid rotor,
\begin{equation}
    \psi(z)=z^p\,\prod_{i=1}^{\cal N} (z^2-x_i), \qquad p=0,1,
\end{equation}
where $p=0$ and $p=1$ correspond to even and odd sectors, respectively. 

Thus, again the cubic algebra framework gives closed-form expressions for the energy spectrum and wave functions in the even and odd sectors of the model in a unified way. As examples, in the following, we present explicit results for the ${\cal N}=0, 1$ cases in both sectors. 

For ${\cal N}=0$, the energy and wavefunction are given by $\frac{a+b}{2}\,j(j+1)+\frac{2c-a-b}{2}\,(2{\cal N}+p-j)^2$ and $\psi(z)=z^p$, respectively. Here $p$ (which takes the value 0 or 1) and the spin $j$ are constrained so that ${\cal N}=j-(p+q)/2$, where $q=0,1$, is equal to 0. In this case, the energies and wavefunctions in the even sectors ($p=0$) are given by
\begin{equation}
    E_{even}=0,\qquad \psi_{even}(z)=1,
\end{equation}
\begin{equation}
 E_{even}=\frac{a+b+c}{4},\qquad \psi_{even}(z)=1, 
\end{equation}   
which correspond to $j=0$ (integer) and $j=1/2$ (half-integer), respectively.
The energies and wavefunctions in the odd sector ($p=1$) are 
\begin{equation}
    E_{odd}=a+b,\qquad \psi_{odd}(z)=z,
\end{equation}
\begin{equation}
 E_{odd}=\frac{a+b+c}{4},\qquad \psi_{odd}(z)=z, 
\end{equation}
associated with $j=1$ (integer) and $j=1/2$ (half-integer), respectively.

For ${\cal N}=1$, there are two sets of energies and wavefunctions in the the even sector ($p=0$),
\begin{equation}
\begin{split}
        &E_{even}=a+c,\qquad \psi_{even}(z)=z^2+1,\\
        &E_{even}=b+c,\qquad \psi_{even}(z)=z^2-1.
\end{split}
\end{equation}
and
\begin{equation}
\begin{split}
     &E_{even}=\frac{5(a+b+c)}{4}\mp \sqrt{a^2+b^2+c^2-ab-bc-ac},\\
     &\psi_{even}(z)=z^2-\frac{a+b-2c \pm 2\sqrt{a^2+b^2+c^2-ac-bc-ab}}{3(a-b)}, 
\end{split}
\end{equation}
which correspond to $j=1$ (integer) and $j=3/2$ (half-integer), respectively.
Similarly, the two sets of energies and wavefunctions in the odd sector ($p=1$) are given by
\begin{equation}
   \begin{split}
       &E_{odd}=4a+b+c,\qquad \psi_{odd}(z)=z^3+ z,\\
       &E_{odd}=a+4b+c,\qquad \psi_{odd}(z)=z^3- z
   \end{split} 
\end{equation}
and
\begin{equation}
\begin{split}
     &E_{odd}=\frac{5(a+b+c)}{4}\mp\sqrt{a^2+b^2+c^2-ab-bc-ac},\\
     &\psi_{odd}(z)=z^3+\frac{a+b-2c \pm 2\sqrt{a^2+b^2+c^2-ac-bc-ab}}{a-b}z.
\end{split}
\end{equation}
associated with $j=2$ (integer) and $j=3/2$ (half-integer), respectively. (Note that other energies and wavefunctions for $j=2$ are obtained from ${\cal N}=2$ solutions.)

\subsubsection{Numerical analysis}

The Hamiltonian of the model (\ref{Asymmetric rotor hamiltonian}) has three parameters $a,b$ and $c$. In this subsection, we provide a numerical analysis of Bethe roots, energy spectrum, and fidelity to describe how they change with the change in model parameters for higher values of ${\cal N}$.

\begin{figure}[h]
  \centering
  \begin{subfigure}{0.25\linewidth}
    \centering
    \includegraphics[width=\linewidth]{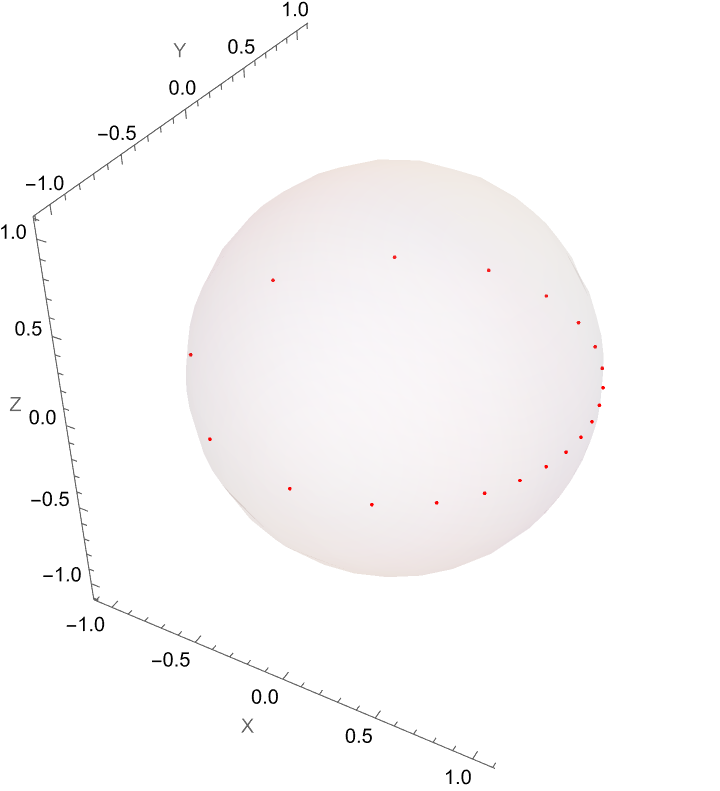}
    \caption{Root distribution for the lowest energy state.}
    \label{f c=-2.4 gs}
  \end{subfigure}
  \hfill
  \begin{subfigure}{0.25\linewidth}
    \centering
    \includegraphics[width=\linewidth]{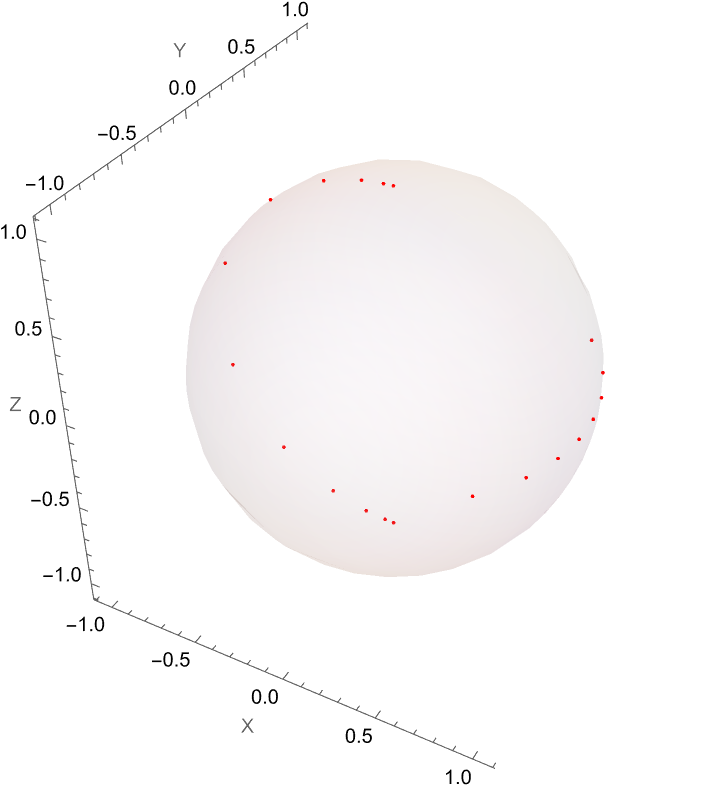}
    \caption{Roots start to gather in the negative part of the real axis.}
    \label{f c=-2.4 s2}
  \end{subfigure}
    \hfill
  \begin{subfigure}{0.25\linewidth}
    \centering
    \includegraphics[width=\linewidth]{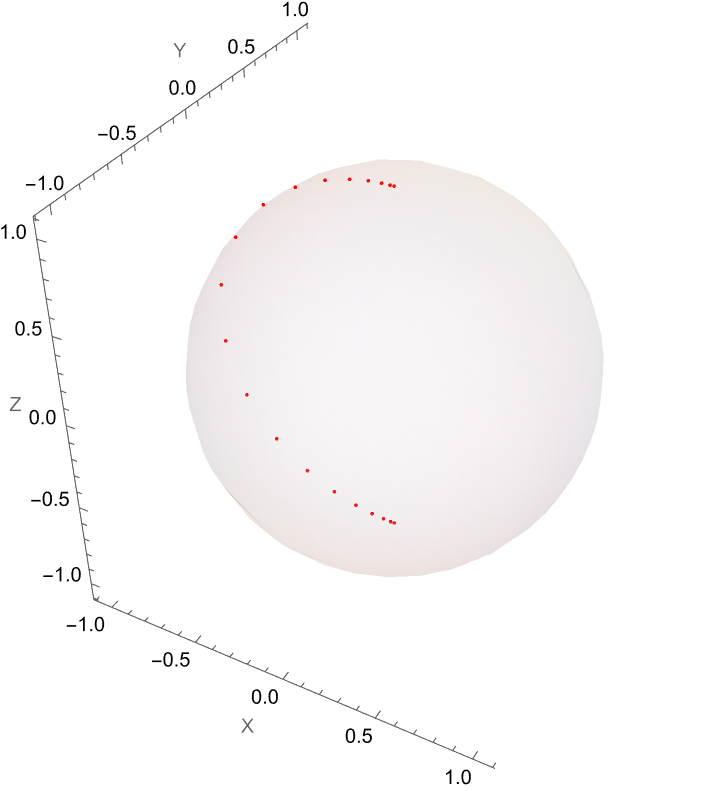}
    \caption{All the roots now lie on the negative part of the real axis.}
    \label{f c=-2.4 s3}
  \end{subfigure}
  \caption{Majorana representation for $c=-2.4$ and change in root pattern with the change of model parameter $c$.}
  \label{fig c=-2.4}
\end{figure}

\begin{figure}[h]
  \centering
  \begin{subfigure}{0.25\linewidth}
    \centering
    \includegraphics[width=\linewidth]{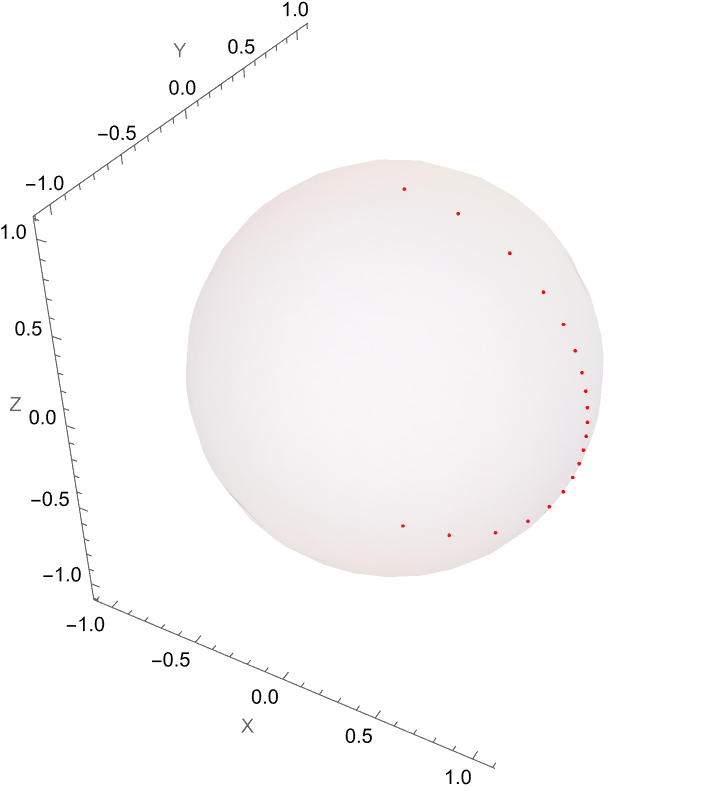}
    \caption{Root distribution for the lowest energy state.}
    \label{f c=2.8 gs}
  \end{subfigure}
  \hfill
  \begin{subfigure}{0.25\linewidth}
    \centering
    \includegraphics[width=\linewidth]{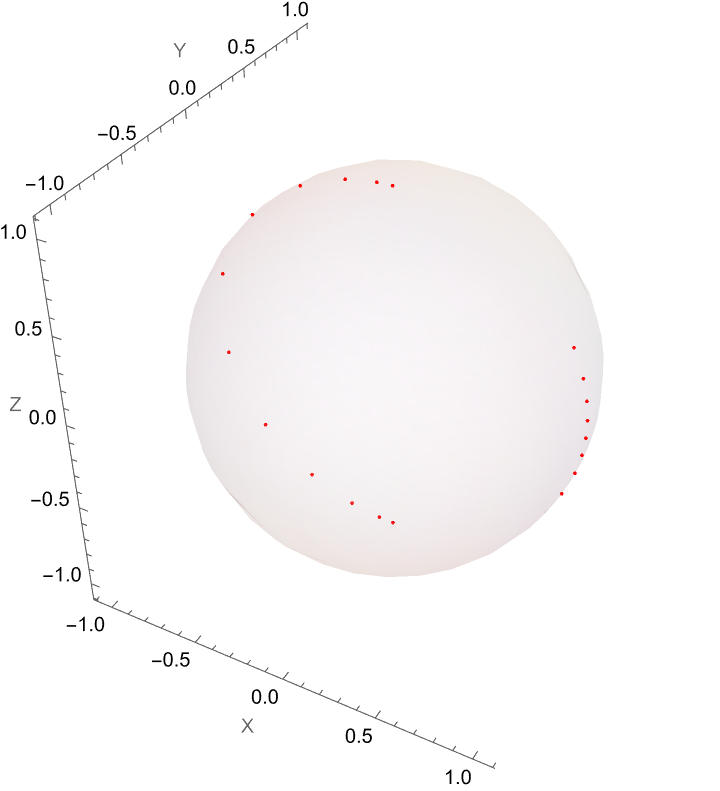}
    \caption{Roots start to gather in the negative part of the real axis.}
    \label{f c=2.8 s2}
  \end{subfigure}
    \hfill
  \begin{subfigure}{0.25\linewidth}
    \centering
    \includegraphics[width=\linewidth]{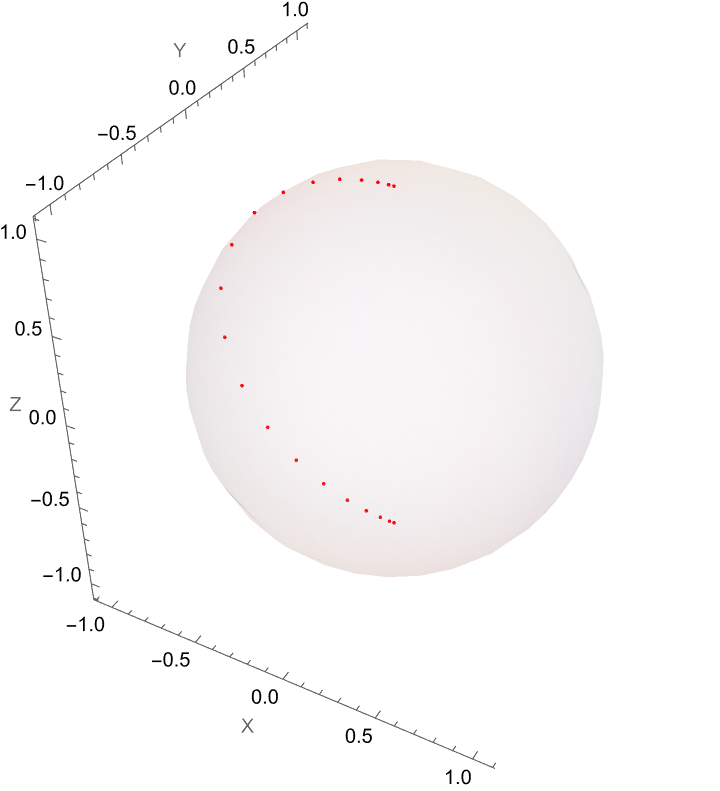}
    \caption{All the roots lie on the negative part of the real axis.}
    \label{f c=2.8 s3}
  \end{subfigure}
  \caption{Majorana representation for $c=2.8$ and change in root pattern with the change in model parameter $c$.}
  \label{fig c=2.8}
\end{figure}

We focus on the case of $j=20={\cal N}$ and set $a=20, b=1.5$ while leaving $c$ as a free parameter. Fig.\ref{fig c=-2.4} and Fig.\ref{fig c=2.8} below are the plots of the distributions of the roots on the unit sphere corresponding to different $c$ values.

\newpage
\noindent Fig.\ref{fig c=-2.4} presents the plots for $c=-2.4$. Fig.\ref{f c=-2.4 gs} is for the lowest energy state. All roots spread on the unit circle in the complex plane. The roots of the excited state are shown in Fig.\ref{f c=-2.4 s2}.  Fig.\ref{f c=-2.4 s3} shows that all roots are on the negative side of the real axis. When the parameter $c$ reaches the value 2.8, the root distribution changes for the lowest energy state. 
Fig.\ref{fig c=2.8} shows the plots for $c=2.8$. As seen in Fig.\ref{f c=2.8 gs}, the roots of the lowest energy state are distributed on the positive side of the real axis in the complex plane. The roots move to the negative side of the real axis when the energy is increased, as shown in Fig.\ref{f c=2.8 s2} and Fig.\ref{f c=2.8 s3}. The difference in the patterns of the roots between Fig.\ref{f c=-2.4 gs} and Fig.\ref{f c=2.8 gs} is less visible, compared with the corresponding results of the LMG model. 

We also calculate and plot the energy spectrum in the odd and even sectors in Fig.\ref{f asymmetry Top energy} below. We can observe that the even and odd energies are non-degenerate for $c\leq 1.5$ and degenerate when $c>1.5$. However, this change in degeneracy only appears for some lower-energy states. Fig.\ref{f asymmetry Top Energy zoom} presents a section of the plot for the energy with $c$ close to 2. The change in degeneracy of the energies at $c=1.5$ is evidence of different regions and a possible phase transition at that point.  

\begin{figure}[h]
  \centering
  \begin{subfigure}{0.56\linewidth}
    \centering
    \includegraphics[width=\linewidth]{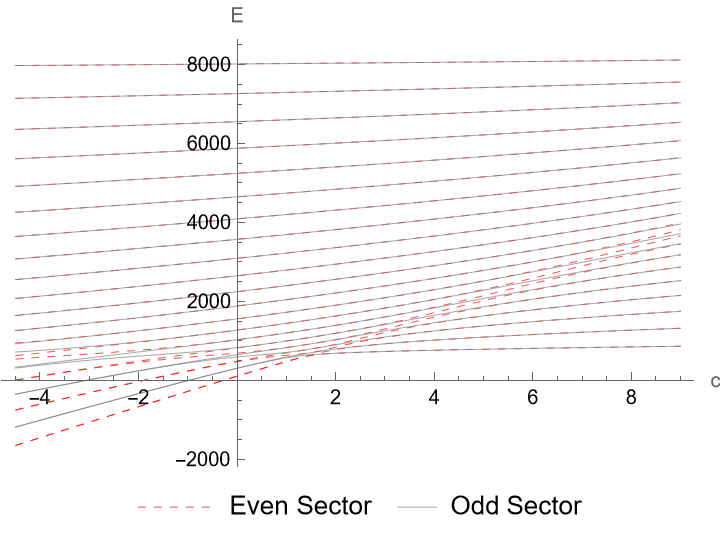}
    \caption{Energies in the even and odd sectors. }
    \label{f asymmetry Top energy}
  \end{subfigure}
  \hfill
  \begin{subfigure}{0.36\linewidth}
    \centering
    \includegraphics[width=\linewidth]{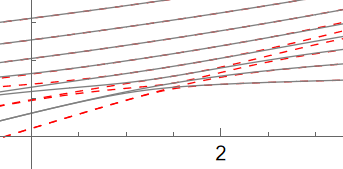}
    \caption{Zoom-in of a section of the energies in Fig.7(a).}
    \label{f asymmetry Top Energy zoom}
  \end{subfigure}
  \caption{Degeneracy of the energy levels in the odd and even sectors.}
  \label{fig asymmetry Top}
\end{figure}

\begin{figure}[h]
  \centering
  \begin{subfigure}{0.32\linewidth}
    \centering
    \includegraphics[width=\linewidth]{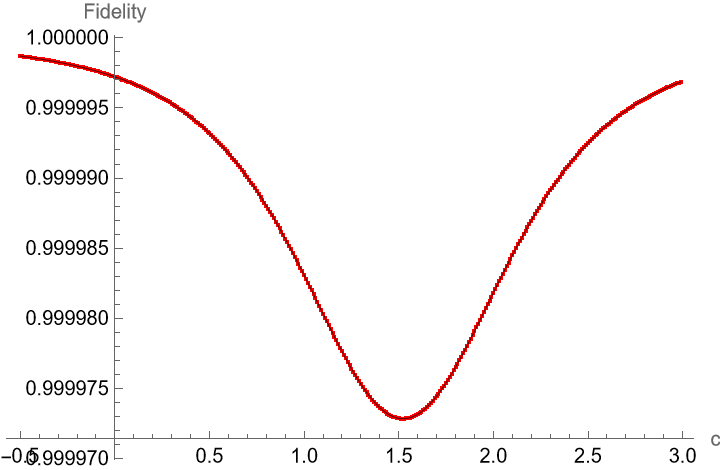}
    \caption{Fidelity of the lowest energy state.}
    \label{f fidelity asymmetry Top Molecule}
  \end{subfigure}
  \hfill
  \begin{subfigure}{0.32\linewidth}
    \centering
    \includegraphics[width=\linewidth]{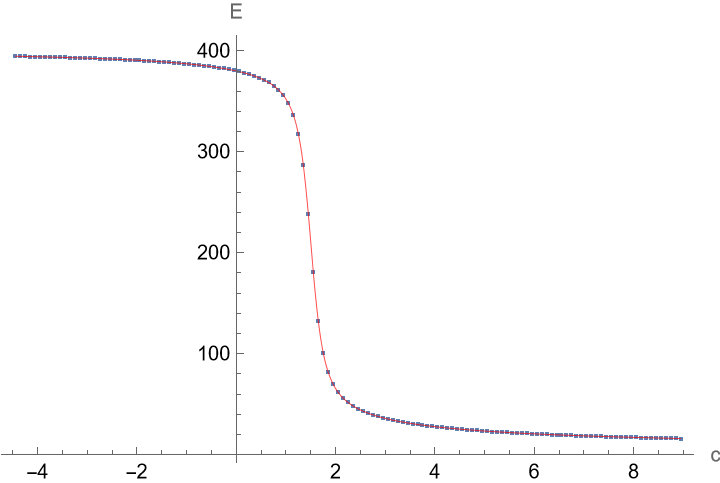}
    \caption{1st order derivative of the energy.}
    \label{f asymmetry Top 1st derivative}
  \end{subfigure}
    \hfill
  \begin{subfigure}{0.32\linewidth}
    \centering
    \includegraphics[width=\linewidth]{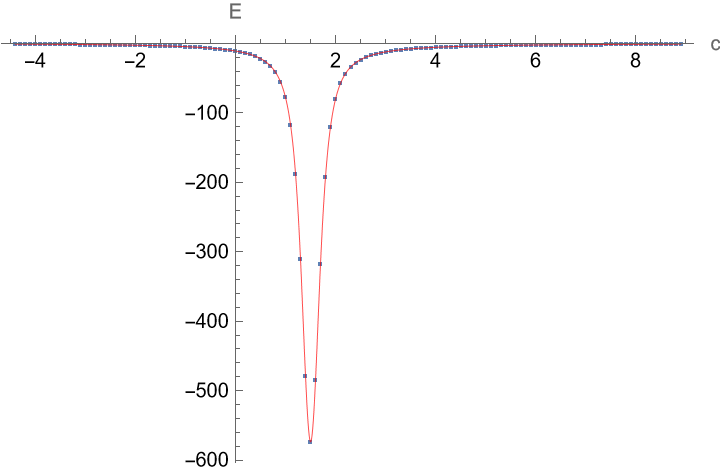}
    \caption{2nd order derivative of the energy.}
    \label{f asymmetry Top 2nd derivative}
  \end{subfigure}
  \caption{Fidelity and derivatives of the energy.}
  \label{fig asymmetry Top others}
\end{figure}

The fidelity and derivatives of the energy are calculated using the numerical Bethe roots and shown in Fig.\ref{fig asymmetry Top others}. Fig.\ref{f fidelity asymmetry Top Molecule} provides the fidelity of the energy. The parameter value related to the minimum value of fidelity is $c=1.5$. This also indicates different regions and a possible phase transition. Fig.\ref{f asymmetry Top 1st derivative} and Fig.\ref{f asymmetry Top 2nd derivative} show how the derivatives of the energy change with the model parameter. A sudden change in the second-order derivative of the energy at $c=1.5$ also indicates a possible phase transition.

\newpage
\subsection{Two-axis countertwisting squeezing model}\label{2-axis}

The Hamiltonian of the two-axis countertwisting squeezing model is 
\begin{equation}\label{Two-axis hamiltonian}
   H= \frac{\chi}{2i} (J_x J_y + J_y J_x)=\frac{\chi}{2i}(J_{+}^2 - J_{-}^2).
\end{equation}
This Hamiltonian has an algebraization in terms of $pl(sl(2))$ of degree 3 corresponding to $k=2$ in (\ref {pl(su(2)) commutation relations}). That is, the hidden-symmetry algebra for this model is a cubic algebra.

\subsubsection{Closed-form solutions}
Similar to the preceding two subsections, we apply the general framework to the two-axis countertwising Hamiltonian (\ref{Two-axis hamiltonian}). We find  $p=0,1$ and
\begin{eqnarray}\label{N for two-axis}
    {\cal N}=j-\frac{p+q}{2}
\end{eqnarray}
with $q=0,1$ so that ${\cal N}$ is a non-negative integer. In terms of the differential realization of the corresponding cubic symmetry algebra, the gauge-transformed Hamiltonian $H$ of the model takes the form of 2nd-order differential operator and the Hamiltonian equation for $H$ is the Schr\"odinger equation which has the form 
\begin{equation}
    (4x^3-4x)\frac{d^2\varphi}{dx^2}+2[(3-4j+2p)x^2-(1+2p)x]\frac{d\varphi}{dx}+\left[(p-2j)(1-2j+p)x-\frac{2iE}{\chi}\right]\varphi=0. 
\end{equation}
Solutions of this equation are polynomials of degree ${\cal N}$
\begin{equation}
    \varphi(x)=\prod_{i=1}^{\cal N} (x-x_i), \qquad \varphi(x)\equiv 1 ~\text{for}~{\cal N}=0,
\end{equation}
where $\{x_i\}$ are the roots of Bethe ansatz equations
\begin{equation}\label{Two-axis BAE}
        \sum_{\ell\neq i}^{\cal N} \frac{2}{x_\ell-x_i}+\frac{(3-4j+2p)x_i^2-(1+2p)}{2( x_i^3+x_i)}=0, \qquad i=1,2,\cdots,{\cal N}. 
\end{equation}
The corresponding energy eigenvalue $E$ is given by
\begin{align}\label{Two-axis energy}
    E=i\frac{\chi}{2}( 2j-p+1-2{\cal N})(2j-p+2-2{\cal N})\,\sum_{i=1}^{\cal N} x_i
\end{align}
in terms of the roots of the Bethe ansatz equations. To our knowledge, general closed-form expression for the energy eigenvalues of the two-axis countertwisting model has not been obtained previously.

We remark that as can be seen from the exact results below for ${\cal N}=0, 1$ and from numerical simulations corresponding to large ${\cal N}$ in next subsection, the roots $\{x_i\}$ of the Bethe ansatz equations (\ref{Two-axis BAE}) are imaginary.  That is, the energy spectrum (\ref{Two-axis energy}) of the two-axis countertwising model is real, as expected from the hermiticity of the Hamiltonian. 

Similar to the treatment for the LMG and the asymmetric rotor models in the preceeding subsections, from the constraint (\ref{N for two-axis}), i.e.
\begin{equation}
    j={\cal N}+\frac{p+q}{2},\quad {\cal N}\in\mathbb{Z}_+; \quad p=0,1;\quad q=0,1, 
\end{equation}
we have the following four cases for the values of $j, p$ and $q$ so that ${\cal N}$ is non-negative integer:
\begin{eqnarray}
    &&j~\text{even}:\quad p=q=0 \quad\text{or}\quad p=1=q,\\
    &&j~\text{odd}:\quad p=0,~ q=1\quad\text{or}\quad p=1,~ q=0,
\end{eqnarray}
We now include the gauge factor and apply $x=z^2$ to obtain the wavefunction in terms of $z$ for the two-axis countertwisting model,
\begin{equation}
    \psi(z)=z^p\,\prod_{i=1}^{\cal N} (z^2-x_i), \qquad p=0,1,
\end{equation}
where $p=0$ and $p=1$ correspond to even and odd sectors, respectively. 

Thus, again the general framework gives closed-form expressions for the energy spectrum and wave functions in the even and odd sectors of the model in a unified way. As examples, in the following, we present explicit results for the ${\cal N}=0, 1$ cases in both sectors. 

For ${\cal N}=0$, the energy equals zero, i.e. $E=0$ and wavefunction is $\psi(z)=z^p$. Here $p$ (which takes the value 0 or 1). Note that the spin $j$ are constrained so that ${\cal N}=j-(p+q)/2$, where $q=0,1$, is equal to 0. In this case, the energies and wavefunctions in the even sectors ($p=0$) are given by
\begin{equation}
    E_{even}=0,\qquad \psi_{even}(z)=1,
\end{equation}
\begin{equation}
 E_{even}=0,\qquad \psi_{even}(z)=1, 
\end{equation}   
which correspond to $j=0$ (integer) and $j=1/2$ (half-integer), respectively. 
The energies and wavefunctions in the odd sector ($p=1$) are 
\begin{equation}
    E_{odd}=0,\qquad \psi_{odd}(z)=z,
\end{equation}
\begin{equation}
 E_{odd}=0,\qquad \psi_{odd}(z)=z, 
\end{equation}
associated with $j=1$ (integer) and $j=1/2$ (half-integer), respectively.

For ${\cal N}=1$, there are two sets of energies and wavefunctions in the the even sector ($p=0$),
\begin{equation}
    E_{even}=\pm \chi,\qquad \psi_{even}(z)=z^2 \pm i; 
\end{equation}
\begin{equation}
 E_{even}=\pm \sqrt{3}\chi,\qquad \psi_{even}(z)=z^2\pm \frac{\sqrt{3}i}{3},
\end{equation}
which correspond to $j=1$ (integer) and $j=3/2$ (half-integer), respectively.
Similarly, the two sets of energies and wavefunctions in the odd sector ($p=1$) are given by
\begin{equation}
    E_{odd}=\pm 3\chi,\qquad \psi_{odd}(z)=z^3 \pm iz;
\end{equation}
\begin{equation}
 E_{odd}=\pm \sqrt{3}\chi,\qquad \psi_{odd}(z)=z^3\pm i\sqrt{3}z, 
\end{equation}
associated with $j=2$ (integer) and $j=3/2$ (half-integer), respectively. (Note that other energies and wavefunctions for $j=2$ are obtained from ${\cal N}=2$ solutions.

\subsubsection{Numerical analysis}

To show how the Bethe root pattern changes with increasing energy for higher value of ${\cal N}$, in Fig.\ref{fig two axis chi=0.4} and Fig.\ref{fig two axis chi=1} below, we plot the Bethe roots for ${\cal N}=20$ in the even sector with $\chi=0.4$ and $\chi=1$ on a unit sphere, respectively. 

\newpage

\begin{figure}[h]
  \centering
  \begin{subfigure}{0.25\linewidth}
    \centering
    \includegraphics[width=\linewidth]{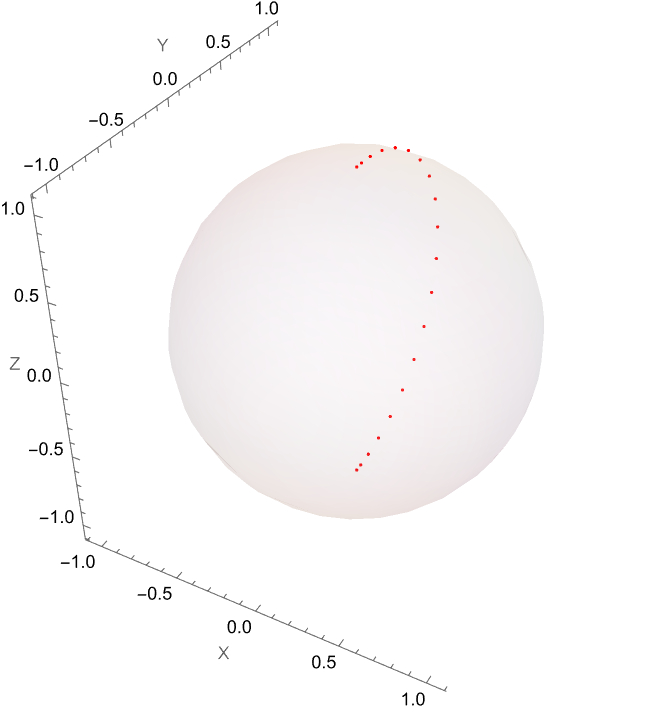}
    \caption{Roots for the lowest energy state.}
    \label{f two axis chi=0.4 gs}
  \end{subfigure}
  \hfill
  \begin{subfigure}{0.25\linewidth}
    \centering
    \includegraphics[width=\linewidth]{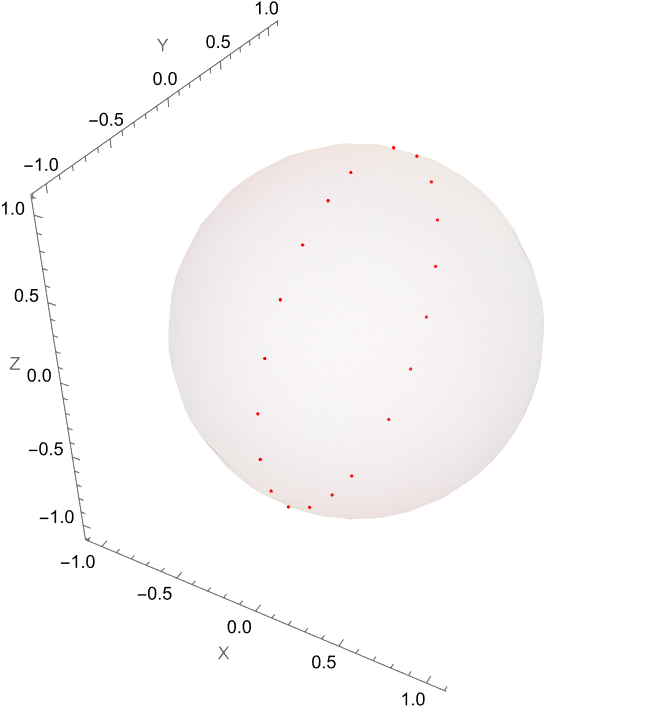}
    \caption{Roots spread around the imaginary axis.}
    \label{f two axis chi=0.4 s2}
  \end{subfigure}
    \hfill
  \begin{subfigure}{0.25\linewidth}
    \centering
    \includegraphics[width=\linewidth]{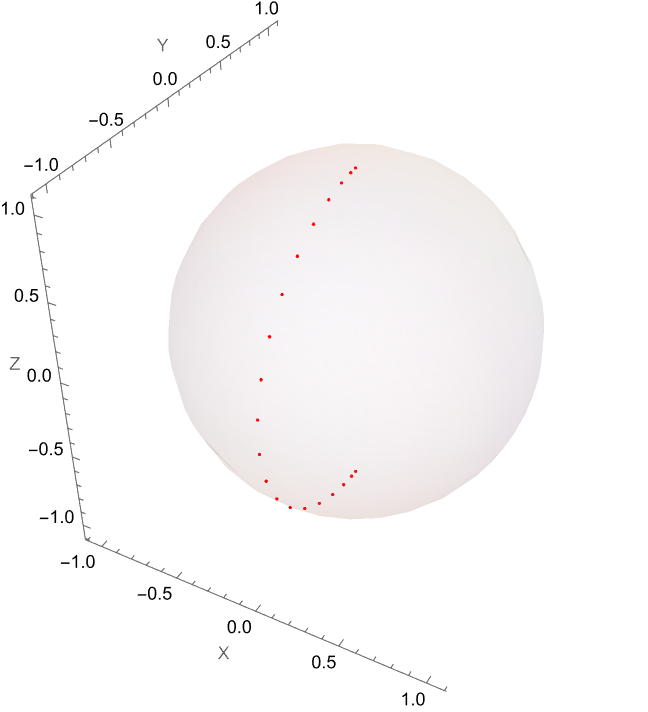}
    \caption{All the roots lie on the negative part of the imaginary axis.}
    \label{f two axis chi=0.4 s3}
  \end{subfigure}
  \caption{Majorana representation for $\chi=0.4$ and change in root distribution.}
  \label{fig two axis chi=0.4}
\end{figure}

\begin{figure}[h]
  \centering
  \begin{subfigure}{0.25\linewidth}
    \centering
    \includegraphics[width=\linewidth]{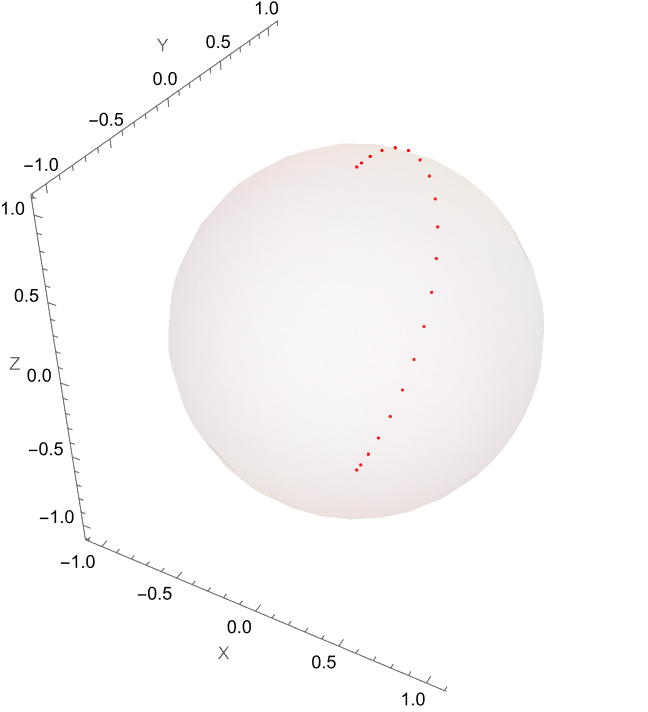}
    \caption{Roots for the lowest energy state.}
    \label{f two axis chi=1 gs}
  \end{subfigure}
  \hfill
  \begin{subfigure}{0.25\linewidth}
    \centering
    \includegraphics[width=\linewidth]{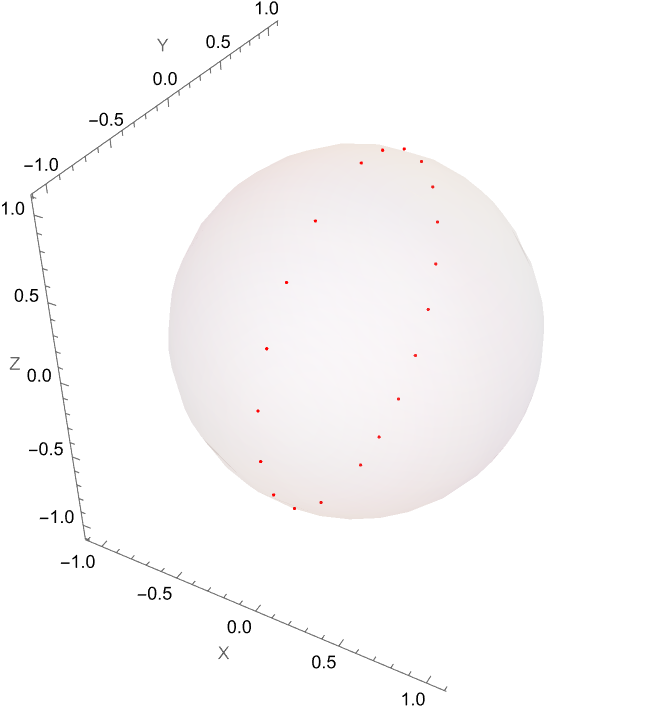}
    \caption{Roots spread around the imaginary axis.}
    \label{f two axis chi=1 s2}
  \end{subfigure}
    \hfill
  \begin{subfigure}{0.25\linewidth}
    \centering
    \includegraphics[width=\linewidth]{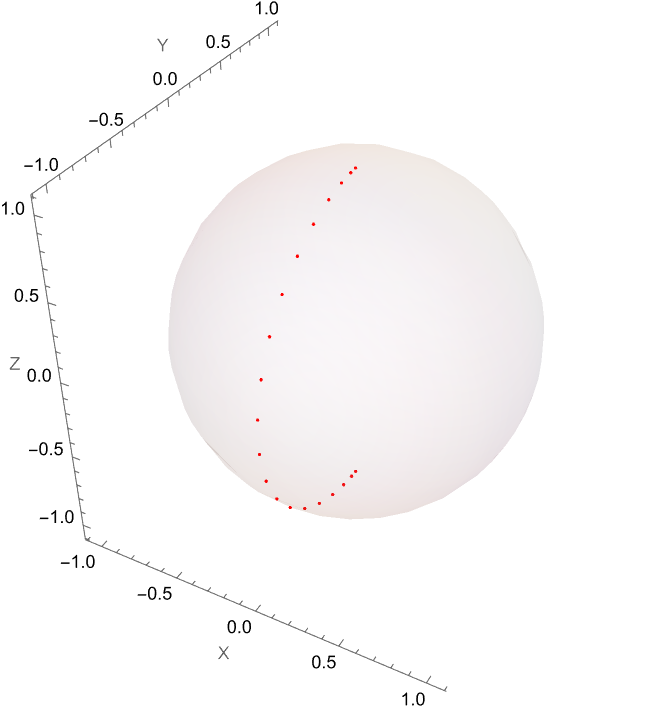}
    \caption{All the roots lie on the negative part of the imaginary axis.}
    \label{f two axis chi=1 s3}
  \end{subfigure}
  \caption{Majorana representation for $\chi=1$ and change in root distribution.}
  \label{fig two axis chi=1}
\end{figure}

As seen in Fig.\ref{f two axis chi=0.4 gs}, the roots of the lowest energy state are located on the positive side of the $y$-axis.  They then move towards the negative direction of the $y$-axis with increasing energy and settle on the negative side of the imaginary axis, as shown in Fig.\ref{f two axis chi=0.4 s2} and Fig.\ref{f two axis chi=0.4 s3}. 
Similar plots for $\chi=1$ are shown in Fig.\ref{fig two axis chi=1}. We find that the corresponding Bethe root distributions are the same as those in Fig.\ref{fig two axis chi=0.4} for $\chi=0.4$.  Note that all roots are purely imaginary so that the corresponding energies are real, as expected from the hermiticity of the Hamiltonian.

\begin{figure}[h]
    \centering
    \includegraphics[width=0.5\linewidth]{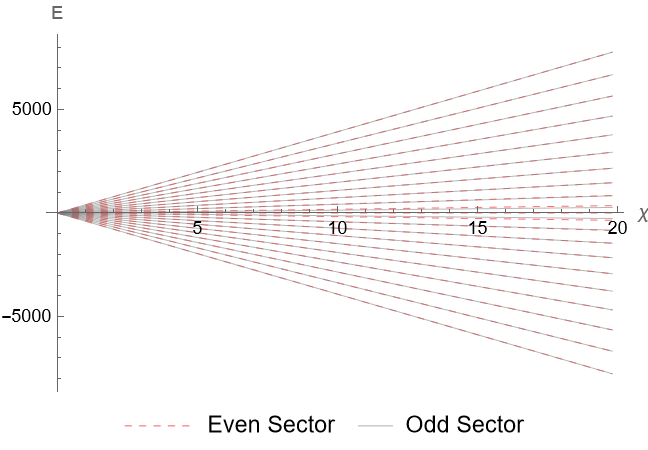}
    \caption{Energy spectrum in the even and odd sectors.}
    \label{fig two axis Energy}
\end{figure}

Unlike the LMG and asymmetric rigid rotor models in the previous subsections,  Fig.\ref{fig two axis chi=0.4} and Fig.\ref{fig two axis chi=1} seem to indicate that the two-axis countertwisting system is always in equilibrium without any phase transition.  
To provide more concrete evidence for this, we also examine the energy spectrum, derivatives of the energy and fidelity of the system for ${\cal N}=20$. Fig.\ref{fig two axis Energy} below is the plot for the energy spectrum in the odd and even sectors with $\chi$ varying from $\chi=0.1$ to $\chi=20$. It shows that the even and odd energies are degenerate. 

\begin{figure}[h]
  \centering
  \begin{subfigure}{0.32\linewidth}
    \centering
    \includegraphics[width=\linewidth]{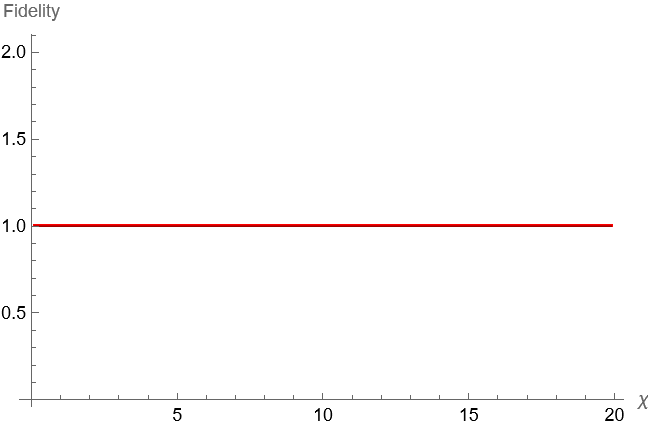}
    \caption{Fidelity of the lowest energy state.}
    \label{f fidelity two axis}
  \end{subfigure}
  \hfill
  \begin{subfigure}{0.32\linewidth}
    \centering
    \includegraphics[width=\linewidth]{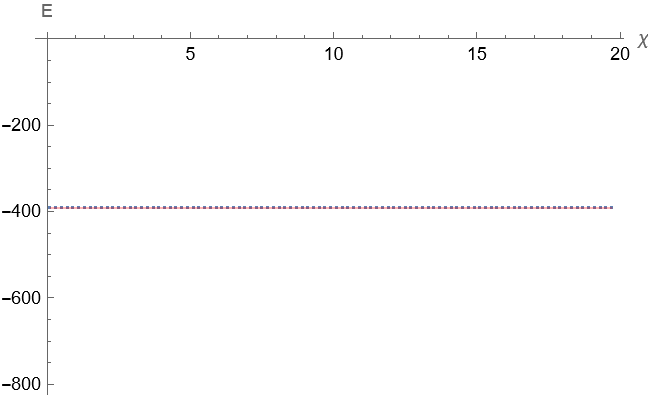}
    \caption{1st order derivative of the energy.}
    \label{f two axis 1st derivative}
  \end{subfigure}
    \hfill
  \begin{subfigure}{0.32\linewidth}
    \centering
    \includegraphics[width=\linewidth]{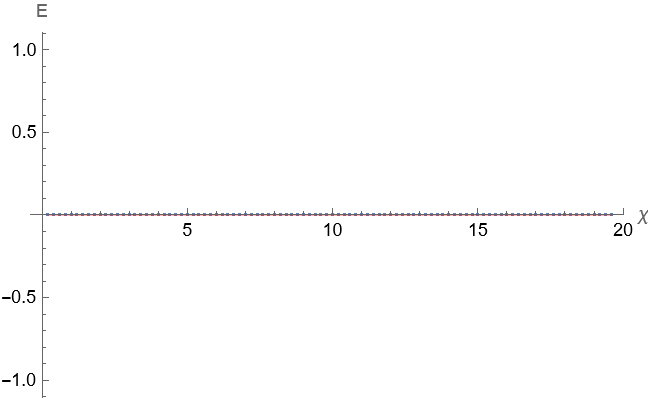}
    \caption{2nd order derivative of the energy.}
    \label{f two axis 2nd derivative}
  \end{subfigure}
  \caption{Fidelity and derivatives of the energy.}
  \label{fig two axis others}
\end{figure}

\newpage
The plots for the the derivatives and fidelity of the lowest energy state calculated by the Bethe roots are shown in Fig.\ref{fig two axis others}. As seen in Fig.\ref{f fidelity two axis}, the fidelity is equal to 1, indicating that the system is always in the same phase region. The change with $\chi$ for 1st- and 2nd-order derivatives of the energy is plotted in Fig.\ref{f two axis 1st derivative} and Fig.\ref{f two axis 2nd derivative}, respectively. They show that the slope of the energy is always the same. These reinforce the above deduction that with the change of model parameter $\chi$ the two-axis countertwisting model is always in the same phase region (i.e. no phase transition). 

%\newpage
\section{Conclusion}\label{Conclusion}

One of the main results of this paper is the development of a systematic algebraic framework for a unified treatment of a class of spin models. We have introduced a novel polynomial algebra and constructed its finite-dimensional representations and the corresponding single-variable differential realization. We have applied our general formalism to three spin models of physical significance: the LMG model, the molecular asymmetric rigid rotor, and the two-axis countertwisting squeezing model. We have shown that all three models have a hidden cubic algebra symmetry in the sense that their respective Hamiltonian admits an algebraization in terms of the generators of their cubic symmetry algebras.  This has allowed us to obtain the closed-form expressions for the energies and wavefunctions of the models via the Bethe ansatz method in a unified manner.
In the Appendix, we have re-derived the exact solutions associated with each of the three spin models by a direct 3-term recursive relation approach. This leads to new types of polynomials in the sense that, unlike the so-called Bender-Dunne polynomials \cite{bender1996}, each of them has two critical polynomials whose zeros give the energy eigenvalues of the corresponding spin model. 

We have provided a numerical analysis on the roots of the Bethe ansatz equations of the three spin models for higher values of ${\cal N}$. To characterize the root distributions, we have displayed their respective Bethe roots on a unit sphere via inverse stereographic projection. This has allowed us to observe the changes in the root patterns of the ${\cal N}=20$ energy states of the models.  Such pattern changes are indications of the appearance of critical points and different phase regions. 
We have also calculated the fidelity and the first- and second-order derivatives of the energy and observed sudden jumps in the plots of these quantities against the model parameters. These results reinforce our claim that possible phase transitions occur in the LMG and asymmetric rigid rotor models.  

The polynomial algebra approach presented in this work provides a powerful framework for analytically solving quasi-exactly solvable systems,  as demonstrated by our unified treatment of the three spin models with cubic algebras as their hidden symmetry algebras. It is expected that our framework can be generalized to other models such as the two-axis two-spin model \cite{Kitzinger2020,pan2021}. Work for this model is in progress and the results will be published elsewhere.

\section*{Acknowledgement}
IM was supported by Australian Research Council Future Fellowship FT180100099, and YZZ was supported by Australian Research Council Discovery Project DP190101529.

\appendix

\section{New types of polynomials associated with the three spin models}

In this appendix, we demonstrate that there is a correspondence between each of the three spin models (with hidden cubic algebra symmetry) and a set of polynomials $\{P_\ell(E)\}$ in the energy $E$  which contain {\em two sets} of critical polynomials $P_{2j+1}(E)$ and $P_{2j+2}(E)$.  In each case,  the zeros of the two sets of critical polynomials are the quasi-exact energy eigenvalues of the model. This is different from the so-called Bender-Dunne polynomials \cite{bender1996} for which there is only one critical polynomial. Such types of polynomials seem new and in our opinion deserve further, systematic investigations.

\subsection{LMG Model}

In terms of the differential operator realization (\ref{differential ops of sl2}), the Hamiltonian (\ref{LMG hamiltonian}) of the LMG model becomes the 2nd-order differential operator,
%\begin{equation}\label{LMG hamiltonian}
%    H=\Delta J_0+g(J_+^2+J_-^2),
%\end{equation}
\begin{equation}\label{LMG H J}
    {\cal H}=g(z^4+1)\frac{d^2}{dz^2}+[-2(2j-1)g z^3+\Delta z]\frac{d}{dz}+2j(2j-1)g z^2-{j \Delta}.
\end{equation}
We seek a solution $\psi(z)$ to the Schr\"odinger equation ${\cal H} \psi(z)=E \psi(z)$ of the form
\begin{equation}\label{LMG series solution}
    \psi(z)=\sum_{\ell=0}^\infty\,P_\ell(E)\,z^\ell.
\end{equation}
Demanding that $\psi(z)$ satisfies the Schr\"odinger equation gives rise to the following 3-term recursion relations for $P_\ell(E)$:
\begin{equation}\label{LMG recursion formula 1}
P_2=\frac{1}{2g}\left(E+ j\,\Delta\right) P_0 ,\qquad P_3=\frac{1}{6g}\left(E + (j-1)\,\Delta\right) P_1,
\end{equation}
\begin{equation}\label{LMG recursion formula 2}
P_{\ell+2} - \frac{E+ (j-l)\Delta}{(\ell+1)(\ell+2) g}P_{\ell} + \frac{(2j+1-\ell)(2j+2-\ell)           }{(\ell+1)(\ell+2)}P_{\ell-2}=0 ,\qquad  \ell \geq 2. 
\end{equation}

It can be shown that the recursion relations generate two sets of monic polynomials which exhibit a significant factorization property. To illustrate this factorization we list in factored form the first few polynomials for the cases of $j=1/2, 1, 3/2$. 

\vskip.1in
\noindent \underline{\bf The $j=1/2$ case}: 
\begin{eqnarray}
  P_2 &=& \frac{1}{2g} \left(E+ \frac{\Delta}{2}\right)P_0,\nonumber\\
  P_3 &=& \frac{1}{6g}\left(E-\frac{\Delta}{2}\right)P_1,\nonumber\\
  P_4 &=&\frac{1}{12g} \left(E- \frac{3}{2}\Delta\right)P_2,\nonumber\\
  P_5&=&\frac{1}{20g} \left(E- \frac{5}{2}\Delta\right) P_3,\nonumber\\
  P_6&=&\frac{1}{360g^2}\left[ \left(E- \frac{7}{2}\Delta\right)\left(E- \frac{3}{2}\Delta\right)-24 g^2\right]P_2.
\end{eqnarray}  
Observe that $P_2(E)$ and $P_3$ are common factors of $P_{2\ell}(E)$ and $P_{2\ell+1}(E)$ for $\ell\geq 2$, respectively. The zeros of $P_2(E)$ and $P_3(E)$ give
\begin{equation}
 E=-\frac{\Delta}{2},\quad P_1=0,\quad \Longrightarrow\quad \psi(z)=P_0,  
\end{equation}
\begin{equation}
 E=\frac{\Delta}{2},\quad P_0=0,\quad \Longrightarrow\quad \psi(z)=P_1 z,   
\end{equation}
which are the two ($=2\times 1/2+1$) exact eigenvalues and the corresponding eigenfunctions for the LMG model with $j=1/2$. 

\vskip.1in
\noindent \underline{\bf The $j=1$ case}: 
\begin{eqnarray}
    P_2&=&\frac{1}{2g}(E+\Delta)P_0,\nonumber\\
    P_3&=&\frac{1}{6g}EP_1,\nonumber\\
    P_4&=&\left[\frac{1}{24g^2}(E-\Delta)(E+\Delta)-\frac{1}{6}\right]P_0,\nonumber\\
    & \vdots &
\end{eqnarray}
We find that in this case $P_3(E)$ and $P_4(E)$ are common factors of $P_{2\ell}(E)$ and $P_{2\ell+1}(E)$ for $\ell\geq 3$, respectively. From the zeros of $P_3(E)$ and $P_4(E)$ we obtain 
\begin{equation}
       E=0,\qquad P_0=0, \quad \Longrightarrow\quad \psi(z)=P_1 z, 
\end{equation}
\begin{equation}
       E=\pm \sqrt{\Delta^2+4g^2}, \quad P_1=0,\quad \Longrightarrow\quad \psi(z)=P_0\left[1+\frac{1}{2g}\left(\Delta \pm \sqrt{\Delta^2+4g^2}\right)z^2\right], 
\end{equation}
which are the three ($=2\time 1+1$) exact eigenvalues and the corresponding eigenfunctions for the LMG model with $j=1$. 

\vskip.1in
\noindent \underline{\bf The $j=3/2$ case}: 
\begin{eqnarray}
    P_2&=&\frac{1}{2g}\left(E+\frac{3}{2}\Delta\right)P_0,\nonumber\\
    P_3&=&\frac{1}{6g}\left(E+\frac{1}{2}\Delta\right)P_1,\nonumber\\
    P_4&=&\left[\frac{1}{24g^2}\left(E-\frac{1}{2}\Delta\right)\left(E+\frac{3}{2}\Delta\right)-\frac{1}{2}\right]P_0,\nonumber\\
    P_5&=&\left[\frac{1}{120g^2}\left(E-\frac{3}{2}\Delta\right)\left(E+\frac{1}{2}\Delta\right)-\frac{1}{10}\right]P_1,\nonumber\\
    & \vdots &
\end{eqnarray}
We find that in this case $P_4(E)$ and $P_5(E)$ are common factors of $P_{2\ell}(E)$ and $P_{2\ell+1}(E)$ for $\ell\geq 4$, respectively. The zeros of $P_4(E)$ and $P_5(E)$ give 
\begin{eqnarray}
E&=& -\frac{1}{2}\Delta \pm \sqrt{\Delta^2+12g^2},\quad P_1=0,\nonumber\\
& &\quad \Longrightarrow\quad \psi(z)=P_0\left[1+\frac{1}{2g}\left(\Delta\pm \sqrt{\Delta^2+12g^2}\right)z^2\right],\\
E&=&\frac{1}{2}\Delta \pm \sqrt{\Delta^2+12g^2}, \quad P_0=0,\nonumber\\ & &\quad \Longrightarrow\quad \psi(z)=P_1\left[ z + \frac{1}{6g}\left(\Delta \pm \sqrt{\Delta^2+12g^2}\right)z^3\right], 
\end{eqnarray}
which are the four ($=2\times 3/2+1$) exact eigenvalues and the corresponding eigenfunctions for the LMG model with $j=3/2$. 

Up to an overall constant, the energy eigenvalues and the corresponding eigenfunctions found above agree with those obtained in subsection 4.1.1 using the cubic algebra approach.  

In general, it can be seen that the factorization occurs because the third term in the recursion relation (\ref{LMG recursion formula 2}) vanishes when $\ell=2j+1$ or $2j+2$, so that all subsequent polynomials $P_{2j+1+2\ell}(E)$ and $P_{2j+1+(2\ell+1)}(E)$ have the common factors $P_{2j+1}(E)$ and $P_{2j+2}(E)$, respectively. In other words, all polynomials $P_{2\ell}(E)$ and $P_{2\ell+1}(E)$ beyond their respective critical polynomials $P_{2j+1}(E)$ and $P_{2j+2}(E)$, respectively factor into a product of of the form
\begin{equation}
    P_{2\ell+2j+1}(E)=P_{2j+1}(E)\,Q_{2\ell}(E),\qquad  P_{(2\ell+1)+2j+1}(E)=P_{2j+2}(E)\,\tilde{Q}_{2\ell}(E),
\end{equation}
where ${Q}_{2\ell}(E)$ and $\tilde{Q}_{2\ell}(E)$ are certain polynomials of degree $2\ell$.
This factorization leads to the result that the zeros of of the critical polynomials $P_{2j+1}(E)$ and $P_{2j+2}(E)$ are the quasi-exact energy eigenvalues because the expansion in (\ref{LMG series solution}) truncates to two sets of polynomials when $E$ is given by the zeros of $P_{2j+1}(E)$ and $P_{2j+2}(E)$.
The two sets of give the even and odd polynomial wavefunctions of the LMG model, respectively. 
These polynomials are called Majorana polynomials.

\subsection{Molecular asymmetry rigid rotor}
 
In terms of the differential operator realization (\ref{differential ops of sl2}), the Hamiltonian (\ref{Asymmetric rotor hamiltonian}) of the molecular asymmetry rigid rotor is expressed as
\begin{equation}
\begin{split}
       {\cal H}=& \frac{a-b}{4} \left[ (z^4 +1) \frac{d^2}{dz^2} -2(2j-1) z^3 \frac{d}{dz} + 2j(2j-1) z^2 \right] \\
       &+ \frac{2c-a-b}{2} \left[ z^2 \frac{d^2}{dz^2} -(2j-1)z \frac{d}{dz} \right] + {j}\left( j c +\frac{a+b}{2}\right). 
\end{split}
\end{equation}
Similar to the procedure in section A.1, we seek a solution $\psi(z)$ to the Schr\"odinger equation ${\cal H} \psi(z)=E \psi(z)$ of the form
\begin{equation}\label{Asymetric rotor series solution}
\psi(z)=\sum_{\ell=0}^\infty\,P_\ell(E)\,z^\ell.
\end{equation}
Demanding that $\psi(z)$ satisfies the Schr\"odinger equation gives rise to the following 3-term recursion relations for $P_\ell(E)$:
\begin{align}
    &\begin{aligned}
        P_2= \frac{1}{2(a-b)} \left[ 4E -2j (2jc+a+b) \right]P_0,
    \end{aligned} \\
    &\begin{aligned}
         P_3= \frac{1}{6(a-b)}\left[  4E -2j (2jc+a+b) +2 (2j-1)(2c-a-b) \right]P_1, 
    \end{aligned} \\
    &\begin{aligned}\label{Recursion relation 3 for rotor}
        & P_{\ell+2} + \frac{1}{a-b} \frac{ 2\ell(\ell-2j)(2c-a-b) +2j (2jc+a+b) -4E}{(\ell+1)(\ell+2)}\,P_\ell\\
        &\qquad\qquad\qquad\qquad\qquad+ \frac{(2j+1-\ell)(2j+2-\ell)}{(\ell+1)(\ell+2)}\,P_{\ell-2} =0,\qquad \ell \geq 2. 
    \end{aligned}
\end{align}

It can be observed that when $\ell=2j+1$ or $2j+2$ the 3rd term in the recursion relation (\ref{Recursion relation 3 for rotor}) vanishes, so that all subsequent polynomials $P_{2j+1+2\ell}(E)$ and $P_{2j+1+(2\ell+1)}(E)$ have common factors $P_{2j+1}(E)$ and $P_{2j+2}(E)$, respectively. Thus factorization occurs and this factorization leads to the result that the zeros of the critical polynomials $P_{2j+1}(E)$ and $P_{2j+2}(E)$ are the quasi-exact energy eigenvalues because the expansion in (\ref{Asymetric rotor series solution}) truncates when the $E$ values are given by the zeros of $P_{2j+1}(E)$ and $P_{2j+2}(E)$.

To illustrate the factorization property we list in factored form the first few polynomials for $j=1/2, 1, 3/2$. 

\vskip.1in
\noindent \underline{\bf The $j=1/2$ case}:
\begin{eqnarray}
    P_2&=&\frac{1}{2(a-b)}\left[4E-(c+a+b)\right]P_0,\nonumber\\
    P_3&=&\frac{1}{6(a-b)}\left[4E-(c+a+b)\right]P_1,\nonumber\\
    P_4&=&\frac{1}{a-b}\left[\frac{1}{4}(a+b-3c)+\frac{E}{3}\right]P_2,\nonumber\\
     P_5&=&\frac{1}{20(a-b)}[11(a+b)-25c+4E]P_3,\nonumber\\
     P_6&=&\left\{\frac{1}{360(a-b)^2}[(3a+3b-9c+4E)(23a+23b-49c+4E)]-\frac{1}{15}\right\}P_2.  
\end{eqnarray}
$P_2$ and $P_3$ are common factors of $P_{2\ell}(E)$ and $P_{2\ell+1}(E)$ for $\ell\geq 2$, respectively. The zeros of $P_2(E)$ and $P_3(E)$ give the eigenvalues. We thus get the energies and wavefunction for $j=1/2$:
\begin{equation}
    E=\frac{a+b+c}{4},\quad P_1=0, \quad \Longrightarrow\quad \psi(z)=P_0,
\end{equation}
\begin{equation}
    E=\frac{a+b+c}{4},\quad P_0=0, \quad \Longrightarrow\quad  \psi(z)=P_1z.
\end{equation}
The energy spectrum for $j=1/2$ is 2-fold degenerate.

\vskip.1in
\noindent \underline{\bf The $j=1$ case}:
\begin{eqnarray}
    P_2&=&\frac{1}{2(a-b)}[4E-2(a+b+2c)]P_0,\nonumber\\
    P_3&=&\frac{2(a+b-E)}{3(a-b)}P_1,\nonumber\\
    P_4&=&\frac{2(a+c-E)(b+c-E)}{3(a-b)^2}P_0,\nonumber\\
    & \vdots &
\end{eqnarray}
We find that in this case $P_3(E)$ and $P_4(E)$ are common factors of $P_{2\ell}(E)$ and $P_{2\ell+1}(E)$ for $\ell\geq 3$, respectively. The zeros of $P_3(E)$ and $P_4(E)$ give rise to the energy eigenvalues and eigenfunctions for $j=1$:
\begin{equation}
    E=a+b,\quad P_0=0, \quad \Longrightarrow\quad \psi(z)=P_1 z. 
\end{equation}
\begin{equation}
    E_1=a+c, \quad P_1=0, \quad \Longrightarrow\quad \psi(z)=P_0\left(1+ z^2\right),
\end{equation}
\begin{equation}
     E_2=b+c,\quad P_1=0, \quad \Longrightarrow\quad  \psi(z)=P_0\left(1-z^2\right). 
\end{equation}

\vskip.1in
\noindent \underline{\bf The $j=3/2$ case}:
\begin{eqnarray}
    P_2&=&\frac{1}{2(a-b)}[4E-3(a+b+3c)]P_0,\nonumber\\
    P_3&=&\frac{1}{6(a-b)}[4E-7a-7b-c]P_1,\nonumber\\
    P_4&=&\left\{\frac{[7(a+b)+c-4E][3(a+b+3c)-4E]}{24(a-b)^2}-\frac{1}{2}\right\}P_0, \nonumber\\
    P_5&=&\left\{\frac{[7(a+b)+c-4E][3(a+b+3c)-4E]}{120(a-b)^2}-\frac{1}{10}\right\}P_1,\nonumber\\
    & \vdots &
\end{eqnarray}
We find that in this case $P_4(E)$ and $P_5(E)$ are common factors of $P_{2\ell}(E)$ and $P_{2\ell+1}(E)$ for $\ell\geq 4$, respectively. From the zeros of $P_4(E)$ and $P_5(E)$ we obtain the energy eigenvalues and eigenfunctions for $j=3/2$:
\begin{equation}
    \begin{split}
        &E=\frac{5(a+b+c)}{4}\pm \sqrt{a^2+b^2+c^2-ab-ac-bc},\quad P_1=0,  \\ 
        &\quad \Longrightarrow\quad\psi(z)=P_0\left[1+\frac{a+b-2c\pm 2\sqrt{a^2+b^2+c^2-ab-bc-ac}}{a-b}\, z^2\right], 
    \end{split}
\end{equation}
\begin{equation}
    \begin{split}
        &E=\frac{5(a+b+c)}{4}\pm \sqrt{a^2+b^2+c^2-ab-ac-bc},\quad P_1=0,\\ 
        &\quad \Longrightarrow\quad\psi(z)=P_1\left[ z-\frac{a+b-2c\pm 2\sqrt{a^2+b^2+c^2-ab-ac-bc}}{3(a-b)}\, z^3\right]. 
    \end{split}
\end{equation}
We remark that the energy spectrum for $j=3/2$ is 2-fold degenerate.

Up to an overall constant, the energy eigenvalues and the corresponding eigenfunctions found above agree with those obtained in subsection 4.2.1 using the cubic algebra approach. 
 
\subsection{Two-axis countertwisting squeezing Hamiltonian}

In terms of the realization (\ref{differential ops of sl2}), the Hamiltonian of the two-axis countertwisting squeezing model becomes the 2nd-order differential operator,
\begin{equation}
 {\cal H}= \frac{\chi}{2i}  \left[ (z^4-1) \frac{d^2}{dz^2} -2(n-1) z^3 \frac{d}{dz} +n(n-1) z^2 \right].    
\end{equation}
we seek a solution $\psi(z)$ to the Schr\"odinger equation ${\cal H} \psi(z)=E \psi(z)$ of the form
\begin{equation}\label{Two-axis series solution}
\psi(z)=\sum_{\ell=0}^\infty\,P_\ell(E)\,z^\ell.
\end{equation}
Demanding that $\psi(z)$ satisfies the Schr\"odinger equation gives rise to the following 3-term recursion relations for $P_\ell(E)$:
\begin{align}
   &\begin{aligned}
         P_2 + \frac{iE}{\chi} P_0 =0 ,\qquad  P_3 + \frac{iE}{3\chi} P_1 =0, 
    \end{aligned}\\
   &\begin{aligned}
      (\ell+1)(\ell+2) P_{\ell+2} + \frac{2iE}{\chi} P_\ell -(2j+1-\ell)(2j+2-\ell) P_{\ell-2} =0 , \qquad \ell \geq 2.    
    \end{aligned}
\end{align}

It can be shown that the critical polynomials for the series solution above are $P_{2j+1}(E)$ and $P_{2j+2}(E)$ and their zeros are precisely the quasi-exact energy eigenvalues of the model.
To illustrate the factorization property we list in factored form the first few polynomials for $j=1/2, 1, 3/2$. 

\vskip.1in
\noindent \underline{\bf The $j=1/2$ case}:
\begin{eqnarray}
    P_2&=&- \frac{iE}{\chi} P_0 =0 ,\qquad  P_3 =- \frac{iE}{3\chi} P_1,\nonumber\\
    P_4&=&-\frac{iE}{6\chi}P_2,\qquad P_5=-\frac{iE}{10\chi} P_3, \nonumber\\
    P_6&=& \frac{1}{15}\left( -\frac{E^2}{6\chi^2} + 1  \right) P_2,\qquad P_7 = \frac{1}{21} \left( - \frac{E^2}{10\chi^2} + 3\right) P_3.
\end{eqnarray}
$P_2(E)$ and $P_3(E)$ are common factors of $P_{2\ell}(E)$ and $P_{2\ell+1}(E)$ for $\ell\geq 2$, respectively. The zeros of $P_2(E)$ and $P_3(E)$ give the eigenvalues. We thus get the energies and wavefunction for the $j=1/2$ case:
\begin{equation}
E=0,\quad P_1=0,\quad \Longrightarrow\quad \psi(z) =P_0,
\end{equation}
\begin{equation}
E=0,\quad P_0=0, \quad \Longrightarrow\quad\psi(z) =P_1 z.
\end{equation}
Note that the energy spectrum for $j=1/2$ is 2-fold degenerate.

\vskip.1in
\noindent \underline{\bf The $j=1$ case}:
\begin{eqnarray}
 P_2&=&- \frac{iE}{\chi} P_0 =0 ,\qquad P_3=-\frac{iE}{3\chi}P_1,\nonumber\\
 P_4&=&\frac{1}{6}\left(1-\frac{E^2}{\chi^2}\right)P_0,\quad \ldots,\quad . 
\end{eqnarray}
It is observed that $P_3(E)$ and $P_4(E)$ are common factors of $P_{2\ell}(E)$ and $P_{2\ell+1}(E)$ for $\ell\geq 3$, respectively. The zeros of $P_3(E)$ and $P_4(E)$ give rise to the energy eigenvalues and eigenfunctions for $j=1$:
\begin{equation}
E=0,\quad P_0=0, \quad \Longrightarrow\quad \psi(z) =P_1 z ,  
\end{equation}
\begin{equation}
    E=\pm \chi,\quad P_1=0, \quad \Longrightarrow\quad\psi(z)=P_0\left(1\mp i z^2\right). 
\end{equation}

\vskip.1in
\noindent \underline{\bf The $j=3/2$ case}:
\vskip.1in
Similarly for this case, we find that $P_4(E)$ and $P_5(E)$ are common factors of $P_{2\ell}(E)$ and $P_{2\ell+1}(E)$ for $\ell\geq 4$, respectively, with
\begin{equation}
    P_4=\frac{1}{6}\left(3-\frac{E^2}{\chi^2}\right)P_0, \qquad P_5=\frac{1}{10}\left(1-\frac{E^2}{3\chi^2}\right)P_1. 
\end{equation}
 The zeros of $P_4(E)$ and $P_5(E)$ give rise to the energy eigenvalues and eigenfunctions for $j=3/2$:
\begin{equation}
E= \pm \sqrt{3}\chi,\quad P_1=0, \quad \Longrightarrow\quad\psi(z)=P_0\left(1 \mp \sqrt{3}i z^2\right), 
\end{equation}
\begin{equation}
    E=\pm \sqrt{3}\chi,\quad P_0=0, \quad \Longrightarrow\quad\psi(z)=P_1\left (z \mp \frac{\sqrt{3}i}{3}  z^3\right). 
\end{equation}
The energy spectrum for $j=3/2$ is 2-fold degenerate.

Up to an overall constant, the energy eigenvalues and the corresponding eigenfunctions found above agree with those obtained in subsection 4.3.1 using the cubic algebra approach.

\bibliographystyle{unsrt}
\bibliography{Reference.bib}

\end{document}